\documentstyle[aps,preprint,epsfig,amssymb,amstex]{revtex}

\begin{document}

\draft
\tighten
\author{A. Bernert$^{1}$, P. Thalmeier$^{2}$, P. Fulde$^{1}$}
\address {$^{1}$ {\it Max-Planck-Institut f{\"u}r Physik komplexer Systeme,
D-01187 Dresden, Germany}\\
$^{2}$ {\it Max-Planck-Institut f{\"u}r Chemische Physik fester Stoffe,
D-01187 Dresden, Germany}}
\date{\today}

\title{A microscopic model for the structural transition and spin gap formation
in $\alpha'$-NaV$_2$O$_5$}

\maketitle
\begin{abstract}
We present a microscopic model for $\alpha'$-NaV$_2$O$_5$. Using an extended
Hubbard model for 
the vanadium layers we derive an effective low-energy model consisting of
pseudospin Ising chains and Heisenberg chains coupled to each other. We find a
``spin-Peierls-Ising'' phase transition which causes charge ordering on every
second ladder and 
superexchange alternation on the other ladders. This transition can be
identified with the first transition of the two closeby transitions 
observed in experiment. Due to charge ordering the effective coupling between
the lattice and the superexchange 
is enhanced. This is demonstrated within a Slater-Koster
approximation. It leads to a second instability with 
superexchange alternation on the charge-ordered ladders due to an alternating 
shift of the O sites on the rungs of that ladder. We can explain within
our model the
observed spin gap, the anomalous BCS ratio, and the anomalous shift of the
critical temperature of the first transition in a magnetic field. To test the
calculated superstructure we determine the low-energy magnon dispersion and find
agreement with experiment.
\end{abstract}


\section {Introduction}
\label {sec1}

The layered oxide $\alpha$'-NaV$_{2}$O$_{5}$ has attracted great interest
since 1996, when Isobe and Ueda reported a phase transition at $T=34$ K with a
spin-Peierls like spin gap formation~\cite {isobe96}. At low
temperatures the spin gap has a size of about $\Delta\approx 100$
K~\cite{isobe96,yosihama98,grenier00}, which yields a BCS
ratio $2\Delta/k_BT_{C}\approx 6$, 
much higher than for other organic or inorganic spin-Peierls
materials, for which it lies around the canonical BCS-value of 3.5. 

Furthermore, experiments have shown 
that there are actually \emph{two} transitions, which
lie very close to each other~\cite{koppen98,fagot00}. Both are of second
order. The first one at
$T_{C1}\approx 34$ K is accompanied by a logarithmic
peak in the specific heat~\cite{powell98,dischnerpriv} while the second at
$T_{C1}-T_{C2}\approx 0.3$ K is of mean-field character
evident from a jump in the specific
heat. NMR measurements suggest that the first transition leads to charge
ordering while the second one opens a spin gap~\cite{fagot00}.

Measurements of the critical exponents yield $\beta_{\delta}\approx
0.15\ldots 0.2$ for the critical exponent of
the lattice distortion $\delta$ 
and $\beta_\Delta\approx 0.34$ for the critical exponent of the spin gap
$\Delta$~\cite{fertey98,ravy99,nakao99,gaulin00}. From these values
the existence of two transition can also be inferred indirectly.
Close to
the critical point the spin gap $\Delta$ due to a lattice distortion $\delta$
is expected to obey $\Delta\propto \delta^{3/4}$~\cite{barnes98},
corresponding to $\beta_{\Delta}=3/4\beta_{\delta}$. This relation is not
fulfilled in $\alpha'$-NaV$_2$O$_5$, indicating the existence of two separate
transitions.

For $T>T_{C1}$ $\alpha'$-NaV$_2$O$_5$ has equivalent V
sites~\cite{schnering98,meetsma98,smolinski98} implying
valence $4.5+$. For $T<T_{C2}$ $^{51}$V-NMR measurements
so far show only two inequivalent sites~\cite {ohama98} while 
x-ray structure determination reports three inequivalent
sites~\cite{ludecke99,deboer00,bernert00}.  

An interesting experimental observation is the magnetic field dependence of
$T_{C1}$. It is only about 25\% of the value expected for a
spin-Peierls transition~\cite{koppen98,schnelle99,bompadre00} 
but on the other hand, it is much higher than
what would be generally expected for a structural
transition~\cite{bulaevskii78}.  

In this article we explain a number of these above mentioned 
features by an analysis of a
microscopic model. We find that the phase transition at $T_{C1}$ can be
regarded as a combination of charge-ordering on every second ladder and
superexchange alternation on the other ladders,
a ``spin-Peierls-Ising'' transition. This also explains the anomalous
shift of $T_{C1}$ in a magnetic field. The phase
transition at $T_{C2}$ 
can be regarded as a spin-Peierls transition on the charge-ordered
ladders. 
It is driven by the charge ordering. The lattice distortion accompanying the
charge ordering increases the coupling constant of a lattice mode invoking
superexchange alternation on the charge-ordered ladders. With
increasing charge ordering for decreasing temperatures the coupling constant 
increases until the second phase transition takes place at $T_{C2}$. 
This second phase transition opens a spin gap in agreement with experimental
observations. Charge ordering is not yet complete at $T_{C2}$. The coupling
constant therefore continues to increase and due to this the low-temperature
spin gap is larger than what would be expected from $T_{C2}$, 
i.e., the BCS ratio is enhanced.

We test our low-temperature structure against x-ray structure determination
and find it to agree within experimental resolution. Furthermore we calculate
the low-energy magnon dispersion and find it to agree nicely with experimental
results from inelastic neutron scattering~\cite{yosihama98,grenier00}. 

This article is organised as follows. In the next section we consider a
microscopic model which we project onto 
a spin-pseudospin model
describing low-energy charge and spin degrees of freedom similar
to~\cite{thalmeier98,sa00,mostovoy00}. We find that a ``zig-zag'' charge
ordering as observed in experiment requires the inclusion of spin and lattice
degrees of freedom, not only the charge degrees of freedom.
We therefore extend the spin-pseudospin
Hamiltonian to include lattice degrees of freedom. 

In section~\ref{sec3} we
analyse this Hamiltonian using RPA. 
We find two phase transitions occuring one
after the other, one being a
combination of charge ordering and spin-dimerisation on different ladders, 
the other being a pure
spin-Peierls transition.

In section~\ref{sec4} we compare the calculated low
temperature structure with experiment. As regards x-ray structure
determination we argue that within experimental resolution theory and
experiment agree. With the calculated low-temperature 
structure we explain the magnon dispersion observed in
experiment. Finally we summarize and discuss our results.

\section {Derivation of the Model Hamiltonian}
\label{sec2}

To construct a model Hamiltonian we note that according to LDA+U
calculations~\cite{yaresko00} the orbitals around the Fermi level are mostly of
vanadium $d_{xy}$ character. They are separated both from the lower lying
oxygen $p$ orbitals as well as from the remaining vanadium $d$-orbitals 
and the sodium 3s orbital which lie energetically higher. This means that
we have to consider only
the quarter filled $d_{xy}$ orbitals. For a discussion of charge
ordering we restrict ourselves to a 2D model neglecting 
hopping matrix elements
and Coulomb interactions between the vanadium layers. The former have been
shown to be negligible both by LDA+U~\cite {yaresko00} and Slater-Koster type
calculations~\cite{bernert00}. The Coulomb interactions 
are expected to be small due to the large 
distance between the V sites of neighbouring layers. They are also
screened by
intermediate O ions. Therefore our inital Hamiltonian represents an
extended one-band Hubbard model at quarter filling, taking 
on-site and intersite Coulomb interactions into account: 
\begin {eqnarray} \label {eq2.1}
H &=& 
\sum_{\ll i,j\gg_R}t_R\left(a_{i\sigma}^\dag a_{j\sigma}+\mathrm{h.c.}\right) 
+\sum_{\ll i,j\gg_L}t_L\left(a_{i\sigma}^\dag a_{j\sigma}+\mathrm{h.c.}\right)
\nonumber \\
&&
+\sum_{\langle i,j\rangle_{IL}}t_{IL}\left(a_{i\sigma}^\dag
a_{j\sigma}+\mathrm{h.c.}\right) 
+\sum_{\ll i,j\gg_D}t_D\left(a_{i\sigma}^\dag
a_{j\sigma}+\mathrm{h.c.}\right) \nonumber \\
&&
+\sum_{\ll i,j\gg_R} V_R n_i n_j 
+\sum_{\ll i,j\gg_L} V_L n_i n_j  \nonumber \\
&&
+\sum_{\langle i,j\rangle_{IL}} V_{IL} n_i n_j
+\sum_{i} U n_{i\uparrow}n_{i\downarrow}.
\end{eqnarray}

\begin {figure}
\epsfig{file=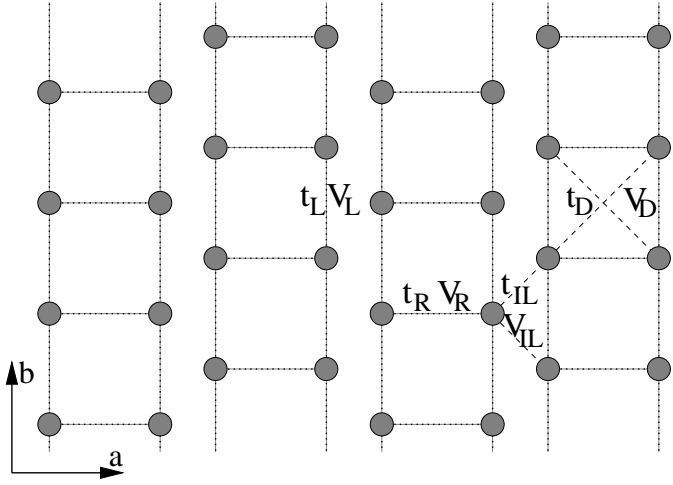}
\vspace{0.25cm}
\caption {Hopping matrix elements and Coulomb interactions in the V layers}
\label {fig1}
\end{figure}

Here $\langle,\rangle_{IL}$ denotes pairs of nearest
neighbour vanadium $d_{xy}$-orbitals while
$\ll,\gg_{R}$ and $\ll,\gg_{L}$ denote pairs of next nearest neighbour 
vanadium $d_{xy}$-orbitals along the rung (R) or the leg (L). 
The first four terms describe the effective hopping of the $d$-electrons
between V sites $i$ and $j$. 
This hopping can take place both via oxygen orbitals and
sodium orbitals. The last four terms describe the intersite and
on-site Coulomb repulsion. These terms connect sites as shown in
Fig.~\ref{fig1}. 

At $T>T_{C1}$ the system is at quarter filling and the V sites are all
equivalent~\cite{schnering98}. Therefore there is an average of 
one electron per rung. The on-site~\cite{smolinski98,horsch98} and the
intersite 
Coulomb repulsion create a charge transfer gap, causing 
$\alpha'$-NaV$_2$O$_5$ to be an insulator.
Therefore hopping
between rungs takes place only virtually. This enables us to
use an effective Hamilton operator $\tilde{H}_{PP}$ which acts on the Hilbert
subspace $\mathcal{P}$ 
of all states with one electron on each
rung. Let $\mathcal{Q}$ be the Hilbert subspace complementary to $\mathcal{P}$
and $P$ be the 
projector onto $\mathcal{P}$ and $Q$ the projector onto $\mathcal{Q}$. An
eigenstate $|\psi_{PP}^{(0)}\rangle$ of $\tilde{H}_{PP}$ satisfies:
\begin{eqnarray} \label{eq2.2}
\tilde{H}_{PP}|\psi_{PP}^{(0)}\rangle &=&
\left(PHP-PHQ\left(QHQ-E_0\right)^{-1}QHP\right)|\psi_{PP}^{(0)}\rangle
\nonumber \\ 
&=& E_0|\psi_{PP}^{(0)}\rangle.
\end{eqnarray}

To calculate $\tilde{H}_{PP}$ we need to know the energies $E_0$ 
which can be either
determined self-consistently or, as for a Schrieffer-Wolff transformation, set
equal to the eigenenergy $E_{0}^{PP}$ of $PHP$. 
In the following we are interested in 
$\tilde{H}_{PP}$ only to order $O\left(t^2/E\right)$, i.e., to second order in
the hopping.  
In this case we can neglect the hopping terms in $\left(QHQ-E_0\right)$, 
since both $PHQ$ and $QHP$ scale with $t$. 

Within the subspace $\mathcal{P}$ we use the operators
$a_{i\alpha\sigma}^\dag$, 
$a_{i\alpha\sigma}$ for the electrons in the atomic orbitals.
Here $i$ denotes the rung, the pseudospin variable
$\alpha=\pm\frac{1}{2}$
describes whether the electron occupies the left ($-\frac{1}{2}$) or right
($+\frac{1}{2}$) V site of a rung and $\sigma$ denotes the z-component
of the spin. Using these operators we define the conditional creation
operators  
\begin{eqnarray*}
\hat{a}_{i\alpha\sigma}^\dag &=& \left(1-n_{i\bar{\alpha}\uparrow}\right) 
\left(1-n_{i\bar{\alpha}\downarrow}\right)
\left(1-n_{i\alpha\bar{\sigma}}\right) a_{i\alpha\sigma}^\dag
\end{eqnarray*}
similar to the operators used in the $t$-$J$-model~\cite{harris67}. 
We then obtain for $\tilde{H}_{PP}$: 
\begin {eqnarray}\label{eq2.3}
\tilde{H}_{PP} &=& H_S+ \sum_i 2\hat{\tilde{t}}_R^iT_i^x +\sum_{\langle
i,j\rangle_{IL}} \hat{K_{IL}}^{ij} T_i^z T_j^z \nonumber \\
&&+\sum_{\ll i,j\gg_{L}} \left(
\hat{K}_{Lz}^{ij} T_i^z T_j^z + \hat{K}_{Lx}^{ij} T_i^x T_j^x +
\hat{K}_{Ly}^{ij} T_i^y 
T_j^y\right) 
\end{eqnarray}
where the $\hat{K}$, $H_{S}$ and $\hat{\tilde{t}}_R$ contain spin operator
products. 
It turns out to be useful to work with spin and pseudospin operators
\begin{eqnarray*}
\vec{S}_i &=& \frac{1}{2}\sum_{\sigma_1,\sigma_2,\alpha}
\hat{a}_{i\alpha\sigma_1}^\dag \vec{\sigma}_{\sigma_1\sigma_2} 
\hat{a}_{i\alpha\sigma_2}, \\
\vec{T}_i &=& \frac{1}{2}\sum_{\sigma,\alpha_1,\alpha_2}
\hat{a}_{i\alpha_1\sigma}^\dag \vec{\sigma}_{\alpha_1\alpha_2} 
\hat{a}_{i\alpha_2\sigma}.
\end{eqnarray*} 
In equation (\ref{eq2.3})
$\vec{S}_i$ denotes the spin of the electron on the $i$-th rung and the z
component of $\vec{T}_i$ corresponds to $\alpha$, so the pseudospin $\vec{T}_i$
describes the hopping of the electron between the two V sites of a
rung. Furthermore
\begin {eqnarray} \label {eq2.4}
H_S &=&
\sum_{\langle i,j\rangle_{L}}
\left(\frac{2t_D^2}{U+\Delta^U_D-E_{0}^{PP}}
+\frac{2t_L^2}{U+\Delta^U_L-E_{0}^{PP}}\right)
\left(\vec{S}_i\vec{S}_j-\frac{1}{4}\right) 
\nonumber \\ 
\hat{\tilde{t}}_{R}^i&=& t_R + \sum_{\langle i,j\rangle_{L}}
t_Lt_D\left(\frac{4\vec{S}_i\vec{S}_j-1} 
{2\left(U+\Delta^U_{LD}-E_{0}^{PP}\right)}\right. \nonumber\\
&& \qquad\qquad\qquad\qquad\left.+ \frac{4\vec{S}_i\vec{S}_j-1}
{2\left(V_R+\Delta^V_{LD}-E_{0}^{PP}\right)}\right) \nonumber \\
\hat{K}_{IL}^{ij}&=&-V_{IL}-\frac{t_{IL}^2\left(4\vec{S}_i\vec{S_j}-1\right)}
{U+\Delta^U_{IL}-E_{0}^{PP}}-\frac{2t_{IL}^2} {V_R+\Delta^V_{IL}-E_{0}^{PP}} \nonumber\\
\hat{K}_{Lz}^{ij} &=& 2V_L + \frac{2t_L^2 \left(4\vec{S}_i\vec{S}_j-1\right)}
{U+\Delta^U_{L}-E_{0}^{PP}} -\frac{2t_D^2 \left(4\vec{S}_i\vec{S}_j-1\right)}
{U+\Delta^U_{D}-E_{0}^{PP}} \nonumber \\
&&
\qquad+ \frac{4t_L^2} {V_R+\Delta^V_{L}-E_{0}^{PP}} - \frac{4t_D^2}
{V_R+\Delta^V_{D}-E_{0}^{PP}}\nonumber \\
\hat{K}_{Lx}^{ij} &=&  \frac{2t_L^2 \left(4\vec{S}_i\vec{S}_j+1\right)}
{V_R+\Delta^V_{L}-E_{0}^{PP}} + \frac{2t_D^2 \left(4\vec{S}_i\vec{S}_j+1\right)}
{V_R+\Delta^V_{D}-E_{0}^{PP}}\nonumber\\
\hat{K}_{Ly}^{ij} &=&  \frac{2t_L^2 \left(4\vec{S}_i\vec{S}_j+1\right)}
{V_R+\Delta^V_{L}-E_{0}^{PP}} - \frac{2t_D^2 \left(4\vec{S}_i\vec{S}_j+1\right)}
{V_R+\Delta^V_{D}-E_{0}^{PP}}.
\end{eqnarray}
Here $\Delta^V$ and $\Delta^U$ are additional  
energy differences due to the local change of occupation numbers 
with hoppings. 
$\hat{K}_{Lz}$ and $\hat{K}_{IL}$ describe the most important interactions
between electrons on neighbouring V sites. They drive the system to
local ``zig-zag'' type ordering or local ``in-line'' type ordering of the
V 3d electrons.
Assuming that we have locally complete ``zig-zag'' short-range 
``ordering'' within a
ladder and no correlations between ladders the $\Delta^V$, $\Delta^U$ are
given by 
\begin {eqnarray} \label {eq2.zz}
\begin {array}{llllllllll}
\Delta_L^V &=& V_L,\quad &\Delta^V_D&=&0,\quad &\Delta^V_{IL}&=&2V_L, \\
\Delta^V_{LD} &=& 0, \quad & \Delta^U_{L}&=&-V_L,\quad& \Delta^U_{D}&=&0, \\
\Delta^U_{IL}&=&-V_{IL},\quad &\Delta^U_{LD}&=&-V_L.
\end{array}
\end{eqnarray}
Assuming that we have local ``in-line'' ``ordering'', they are equal to
\begin {eqnarray}
\begin {array}{llllll}
\Delta^V_L&=&V_{IL}-V_L,\quad& \Delta^V_D&=&2V_{IL}-2V_L,\\
\Delta^V_{IL}&=&V_{IL}-2V_L,\quad& \Delta^V_{LD}&=&0, \\  
\Delta^U_{L}&=&-2V_{IL},\quad& \Delta^U_{D}&=&V_{IL},\\
\Delta^U_{IL}&=&-V_{IL},\quad & \Delta^U_{LD}&=&0.
\end{array}
\end{eqnarray}

We now make a mean-field approximation for the spin 
operator products contained in $\hat{K}$ and
$\hat{\tilde{t}}_R$. With this we obtain an effective pseudospin model for the 
charge degrees of freedom. $K=\langle
\hat{K}\rangle$ and $\tilde{t}_R=\langle \hat{\tilde{t}}_R\rangle$ 
become effective pseudospin
coupling constants. 
This approximation 
should be acceptable since the corrections of the pseudospin part due to
spin-spin-interactions are moderate. 

With this approximation $\tilde{H}_{PP}$ becomes the Hamiltonian for
a strongly anisotropic pseudospin
Heisenberg model in an ``external field'' $2\tilde{t}_R$. We calculate the
effective coefficients $\tilde{t}_R$, $K_L$, $K_{IL}$ assuming that
$\langle\vec{S}_i\vec{S}_{i+1}\rangle=-\frac{3}{4}$ along the ladders and 
$\langle\vec{S}_i\vec{S}_{j}\rangle=\frac{1}{4}$ between ladders. This
assumption is based on the LDA+U results for the sign of the intra-ladder and
inter-ladder exchange constants~\cite{yaresko00}. We use 
$\Delta^V$, $\Delta^U$ from (\ref{eq2.zz})
for the case of local ``zig-zag'' ``ordering'' and set $E_{0}^{PP}$ to be
the ground-state energy for $PHP$. $E_{0}^{PP}$ is then given by the
expression for the ground-state energy of the Ising chain in a transverse
field~\cite{pfeuty70} represented by ${t}_R$: 
\begin {equation}
E_0 = 2\left(2t_R\right)\frac{\left|1-\lambda_0\right|}{\pi}
E\left(2\sqrt{\frac{-\lambda_0}{\left(\lambda_0-1\right)^2}}\right)
\end{equation}
with $\lambda_0 = V_L/2t_R$ and $E$ the complete elliptic integral of second
kind. From the energies calculated in~\cite{yaresko00} by LDA+U 
for different configurations one obtains 
\begin {equation}
2V_L-V_{IL}=0.027\mathrm{~eV}.
\end {equation}
To find $V_R$ we assume that it can be obtained from $V_L$ by scaling by $d^3$
with the
different distances $d$ of the V sites. Using the values for the
parameters from Ref.~\cite{bernert00}, 
i.e., $t_R=-0.172$ eV, $t_L=-0.049$ eV, $t_D=-0.062$
eV, $t_{IL}=0.110$ eV, $V_L=2t_R$ and $U=4$ eV, we find

\begin{eqnarray} \label {eq2.7}
\begin {array}{rcl}
\tilde{t}_R &=& -0.190\mathrm{~eV}\\
K_{IL} &=& -0.677\mathrm{~eV}\\
K_{Lz} &=& 0.679\mathrm{~eV}\\
K_{Lx} &=& -0.027\mathrm{~eV}\\
K_{Ly} &=& 0.011\mathrm{~eV}.
\end{array}
\end{eqnarray}

\begin {figure}
\epsfig{file=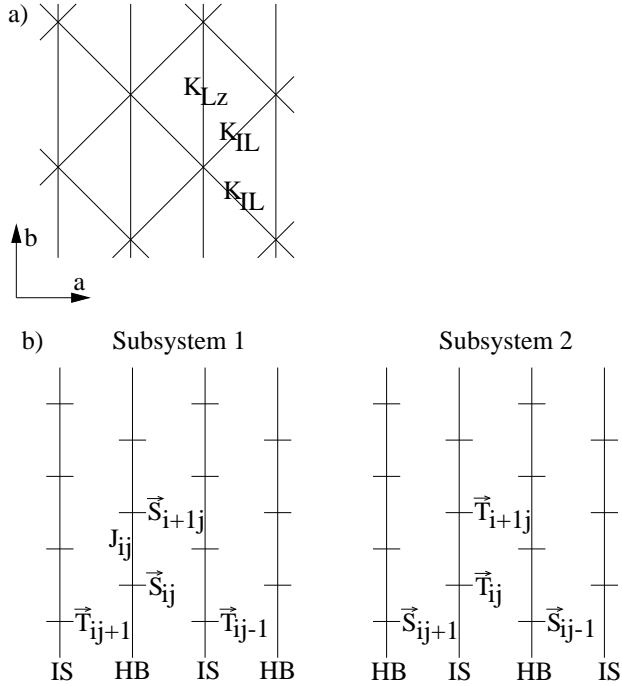}
\vspace{0.25cm}
\caption {a) Lattice geometry if one introduces $\vec{S}_{ij}$ and
$\vec{T}_{ij}$ operators and represents each rung by a point. b) 
The two subsystems of the model described by 
(\ref{eq2.10}) consisting of Ising (IS) pseudospin and Heisenberg (HB) spin 
chains.}
\label {fig24}
\end{figure}

$K_{Lx}$ and $K_{Ly}$ are smaller than the other parameters by more than one
order of magnitude and will therefore be neglected.
Thus our model consists of Ising interactions $K_{Lz}$, $K_{IL}$ and a
``transverse field'' $2\tilde{t}_R$ 
\begin {equation} \label{eq2.6}
H_{1}=\sum_{\ll i,j\gg_L}K_{Lz}T_{i}^zT_{j}^z +\sum_{\langle
i,j\rangle_{IL}}K_{IL}T_{i}^zT_{j}^z +\sum_i 2\tilde{t}_R T_{i}^x.
\end{equation}
The resulting geometry is triangular, as can be seen in
Fig.~\ref{fig24}a. Note that the first term is antiferromagnetic and the
second term is ferromagnetic, resulting in geometrical frustration for the
first two terms of equation (\ref{eq2.7}).

We first consider the case $\tilde{t}_R=0$.
For this case the the relation
between the absolute sizes of $K_{Lz}$ and $K_{IL}$ determines the behaviour
of the system~\cite{houtappel50}. 
If  $|K_{Lz}|<|K_{IL}|$ the system undergoes 
a phase transition at some finite temperature into a
low-temperature 
``ferromagnetic'' state of the pseudospins with 2D-long range order. This state
corresponds to an ``in-line''-ordering of the electrons,
i.e., chains of V$^{4+}$ and V$^{5+}$ alternate along a-direction. 

If  $|K_{Lz}|> |K_{IL}|$ the system 
remains disordered at any finite temperature. At $T=0$
it enters an antiferromagnetic state of the pseudospins with 1D-order along
b-direction 
and no correlation between neighbouring ladders. This corresponds to
a ``zig-zag''-ordering of V$^{4+}$ and V$^{5+}$ along the ladders. Thus the
system is effectively one-dimensional, although there are correlations between
next-nearest-neighbour ladders~\cite{stephenson70}. 

In $\alpha'$-NaV$_2$O$_5$, we have $|K_{Lz}| > |K_{IL}|$. Therefore there is no phase
transition at finite temperatures within the model described by equation
(\ref{eq2.6}), provided that $\tilde{t}_R=0$ as assumed here.
This is consistent with our choice of $\Delta^V$, $\Delta^U$. 
This qualitative
result remains true for other choices for the hopping integrals resulting in
different $K_{Lz}$, $K_{IL}$,
e.g. those obtained from LDA in~\cite{yaresko00}. 
For $\tilde{t}_R=0$ the system is therefore close to a quantum
critical point with a transition from 1D ordering in b-direction to 2D long
range order~\cite{langari01}.

Next we consider the case $\tilde{t}_R\neq 0$. We note that the
total Hamiltonian $H_1$ does not 
imply a phase transition into an ordered state
at any finite temperature. This is due to the fact that 
the first two terms of $H_1$ do not lead to a phase
transition as argued above 
and that the nonzero transverse field suppresses ordering even more.
This result remains qualitatively true if we include
the $K_{Lx}$ and $K_{Ly}$ terms from (\ref{eq2.7})
which also suppress ordering. It also remains true
if we include a small interlayer Ising-like interaction.
This demonstrates that the phase transitions 
observed in $\alpha'$-NaV$_2$O$_5$
cannot be of a purely Coulombic origin: to understand what happens in this
material at low temperatures we have to include at least the spin degrees of
freedom beyond the mean-field approximation used above.
Furthermore, we know from Raman spectroscopy that there is a strong 
coupling between the charge distribution on the V sites and the
distortion of the system~\cite{sherman99}. 
A change in the position of a V sites implies a change in the
positions of the O site on the leg of the neighbouring ladder. The
displacement of the O ions 
can in turn affect the effective superexchange interaction as observed from
experimental data in~\cite{bernert00}. 

Therefore we generalize the Hamiltonian from equation (\ref{eq2.3}), such that
it contains lattice degrees of freedom. Such a Hamiltonian has the following
form: 
\begin {eqnarray} \label {eq2.10}
H_{ISSP} &=& \sum_{i,j} \tilde{K}_{Lz}^{ij}T_{ij}^zT_{i+1j}^z +\sum_{i,j}
2\tilde{t}_R T_{ij}^x +\sum_{i,j} J_{ij}\vec{S}_{ij}\vec{S}_{i+1j} \nonumber
\\
&&+g_{Is}\sum_{ij} T_{ij}^z\vec{e}_0\vec{u}_{ij}
+\sum_{q,j}\omega_{qj}^{VO}b_{VO,qj}^\dag b_{VO,qj} 
 \nonumber \\
&&+\sum_{q,j}\omega_{qj}^{Na}b_{Na,qj}^\dag b_{Na,qj}
+\sum_{q,j}\omega_{qj}^{O_R}b_{O_R,qj}^\dag b_{O_R,qj}. 
\end{eqnarray}
Here $i$ numbers the rungs on a ladder, i.e., it denotes 
the (pseudo)spin on a chain, $j$ denotes the ladder/chain in the lattice and
$\vec{e}_0$ is a unit vector.  
We have introduced phonon operators $b_{qj}^\dag$, $b_{qj}$ for the
displacement 
of neighboring V and O ions (VO) and for the displacement 
of Na or
rung O ions (O$_R$). $\vec{u}$ and $J_{ij}$ are given by
\begin {eqnarray}\label {eq2.11}
J_{ij}&=&\left(1+\vec{u}_{i+\frac{1}{2},j}^{VO}
\vec{\nabla}_{i+\frac{1}{2},j}^{VO}  
+\vec{u}_{i+\frac{1}{2},j}^{Na}\vec{\nabla}_{i+\frac{1}{2},j}^{Na}
+\vec{u}_{i+\frac{1}{2},j}^{O_R}\vec{\nabla}_{i+\frac{1}{2},j}^{O_R}\right) 
\nonumber \\
&&\qquad\qquad\qquad\qquad\qquad\times 
J\left(T_{ij}^z,T_{i+1j}^z\right)\nonumber\\
J\left(T_{ij}^z,T_{i+1j}^z\right) &=&
J_0^{ij}\left(1+f\left(T_{ij}^z,T_{i+1j}^z\right)\right)
\nonumber \\
\vec{u}_{ij}
&=&\sum_{\lambda q}\frac{1} {\left(mN\right)^{1/2}}
\exp\left(i\vec{q}\vec{R}_{ij}\right)Q_j\left(\lambda\vec{q}\right)
\nonumber \\ 
Q_{j}\left(\lambda\vec{q}\right) &=& b_{qj}^\dag + b_{qj} 
.
\end{eqnarray}

$H_{ISSP}$ essentially describes a spin-Peierls model with additional coupling
to a one-dimensional Ising chain in a transverse field. 
It has to take into account the following observations.

At $T=0$ K  there is no correlation between pseudospins of
neighbouring ladders for the pure Ising interaction case. Regarding the
Coulomb interactions we may therefore 
treat the ladders as independent, setting $K_{IL}=0$. The
pseudospin part of the Hamiltonian (\ref{eq2.3}) for each ladder is then
described by a one-dimensional Ising model 
in a transverse field yielding the first two terms of $H_{ISSP}$.
The property that the
Hamiltonian (\ref{eq2.6}) describes a model close to a quantum critical point
has to be included in $H_{ISSP}$. Therefore we have to use an
effective value for $K_{Lz}$ which is close to the value needed for
quantum critical behaviour of the pseudospin Ising chains. This effective
value $\tilde{K}_{Lz}$ is not necessarily identical to the ``true'' value of
this Coulombic coupling in the material.

The spin part of the system behaves like a one-dimensional system at
high temperatures~\cite{isobe96}, i.e., we can use a one-dimensional Heisenberg
model for each ladder to describe it. These terms are contained in the third
term of $H_{ISSP}$. The magnon
dispersion at low-temperatures in a-direction is much smaller than in 
b-direction~\cite{yosihama98,grenier00}, 
even though interladder spin-spin coupling has a
significant value~\cite{yaresko00}. Close to the disordered phase, however,
interladder coupling is reduced due to frustration: as far as interladder
spin-spin-coupling is concerned the system consists of triangles with one
large antiferromagnetic and two smaller ferromagnetic interactions. 

On each ladder spin and pseudospin degrees of freedom are coupled. This
coupling is incorporated into the effective pseudospin coupling $K_{Lz}$ and
the 
effective coupling constant $J_{ij}$ for the superexchange along the ladder. 

A shift of V and neighbouring O ions is assumed to cause an
effective staggered 
field in z-direction for the pseudospins on one leg of a ladder described by
the fourth term of $H_{ISSP}$. Such a shift also causes a
change in the superexchange for the neighbouring leg of the neighbouring
ladder (see Fig.~\ref{fig3}) which has to be incorporated into the 
effective coupling constant $J_{ij}$.

As has been argued in~\cite{bernert00}, a shift of the Na ions in c
direction alternating
along the ladder direction will cause an alternation in the superexchange along
the ladder. The same effect is obtained by an alternating shift of O
ions on a rung in b-direction. Such terms are included in $J_{ij}$ and their 
significance will be examined in section~\ref{sec3.2}.

The last three terms of $H_{ISSP}$ describe the energies of the important 
lattice modes.

The spin-Peierls coupling contained in the expression for $J_{ij}$ in
(\ref{eq2.11}) 
is slightly different from the normal form since it is the shift of the
intermediate O ions or Na ions 
instead of the V ions, which causes
the superexchange dimerisation. 
For parameters we will use
$\tilde{t}_R=-0.190$ eV and $J_{0} = 0.045$ eV for the case 
without charge ordering and no distortion. 

The effective superexchange along a ladder $J_{ij}$ defined in (\ref{eq2.11})
consists of three parts:
$J_0^{ij}$ denotes the original superexchange for the ladder without charge
ordering or distortions. 

The second part, $J_0^{ij}f\left(T_{ij}^z,T_{i+1j}^z\right)$
describes the influence of the charge ordering on the superexchange. From
numerical calculations we know that homogenous charge ordering on a ladder
$j$ causes a drop of the effective superexchange which is quadratic in
$\langle T_j^z\rangle$, as can be seen in Fig.~\ref{fig5}. The parabolic
form of the curve can be understood from noting that the spectral weight of
pair states with two particles on the same rung or the same leg
decreases with $1-(1+\delta_{CO})(1-\delta_{CO})=\delta_{CO}^2$. Analytic
calculations show a similar behaviour~\cite{yushankhai00}. 
Therefore $J_0^{ij}
f\left(T_{ij}^z,T_{i+1j}^z\right)$ is approximately proportional
to $(T_{j}^z)^2$ and the square of the distortion accompanying the charge
ordering. If we use the experimental value for the phonon frequencies
this effect is already contained in the high-temperature values
of the phonon dispersion $\omega_{qj}$ and the term 
$J_0^{ij}f\left(T_{ij}^z,T_{i+1j}^z\right)$ is omitted,
essentially making
a mean-field approximation with regard to the spin operator products for this
expression. 

Finally, there is the essential contribution
\begin {displaymath}
\left(1+\vec{u}_{i+\frac{1}{2},j}^{VO}
\vec{\nabla}_{i+\frac{1}{2},j}^{VO}  
+\vec{u}_{i+\frac{1}{2},j}^{Na}\vec{\nabla}_{i+\frac{1}{2},j}^{Na}
+\vec{u}_{i+\frac{1}{2},j}^{O_R}\vec{\nabla}_{i+\frac{1}{2},j}^{O_R}\right)
J_0^{ij}\left(1+f\left(T_{ij}^z,T_{i+1j}^z\right)\right)
\end{displaymath}
to the superexchange. It
describes the coupling between distortion, superexchange dimerisation and
charge ordering. Here we use a mean-field approximation
for the pseudospin operator
products. The term causes a competition between the spin-Peierls order
parameter and charge order parameter on a ladder.
$\tilde{K}_{Lz}$ will be fitted to properly describe the
system. As argued above, 
we must use $\tilde{K}_{Lz}$ instead of $K_{Lz}$ in equation (\ref{eq2.10}).

\section {Phase transitions}
\label {sec3}

In this section we analyse the Hamiltonian given by equation (\ref{eq2.10}). 
We have two types
of instabilities to consider based on the third and fourth term.
The first type results from 
the fourth term of equation (\ref{eq2.10}) which describes a
coupling between local distortions and charge ordering. By distortion the
system reduces the interaction energy. Due to elastic coupling this is also
accompanied by an alternating exchange coupling on neighbouring ladders.
The second type results from the third term in equation (\ref{eq2.10}) which
includes a coupling between lattice and spin degrees fo freedom
leading to an instability of a spin-Peierls type. 

In the following subsection we
separate the system into two sublattices and first treat each
subsystem on its own. Using RPA we 
analyze in subsection~\ref{sec3.1} the first type of instability 
accompanied by charge ordering and superexchange alternation. We fit the free
parameters $g_{Is}$ and $\tilde{K}_{Lz}$ as to obtain the proper value of
$T_{C1}$ and the shift of $T_{C1}$ in a magnetic field. By analyzing the
ground-state energy of different ordered states we find that at low
temperatures half the ladders are charge-ordered and on the remaining half
there is superexchange alternation.

In subsection~\ref{sec3.2} we examine the influence of charge ordering on
the second type of instability. The distortions of the charge-ordered ladders 
change the coupling
$\vec{u}_{i+\frac{1}{2},j}^{O_R}\vec{\nabla}_{i+\frac{1}{2},j}^{O_R} J_0^{ij}$
and therefore the spin-lattice interaction. It will be shown that this
interaction increases and results in a second phase transition.
As shown below, this explains the anomalous BCS ratio. 

\subsection {Separation of the subsystems}

The Hamiltonian (\ref{eq2.10}) describes a model consisting of two subsystems
1 and 2 shown in Fig.~\ref{fig24}b. Both subsystems contain
Ising pseudospin chains alternating with 
Heisenberg spin chains. The Ising pseudospin chains describe the charge
distribution on a ladder, and the Heisenberg spin chains describe the spin on
a neighbouring ladder. The degrees of freedom on each rung $(ij)$ are
characterised by a pseudospin $\vec{T}_{ij}$ on rung $(ij)$ 
which is contained in one
subsystem and a spin $\vec{S}_{ij}$ contained in the other subsystem. 
The two subsystems 1 and 2 describe the complete
spin-charge dynamics of the Hamiltonian (\ref{eq2.10}). 

Neighbouring chains of the same subsystem are coupled by
the lattice dynamics. A change of the superexchange along a spin chain is
obtained, e.g., by a displacement of an O ion on a leg. Via the  
elastic coupling this causes a displacement
of the neighbouring V site (see Fig.~\ref{fig3}). Such a local
distortion changes the chemical environment and therefore the on-site energy
of the vanadium $d_{xy}$-orbital. This change in energy 
corresponds to a longitudinal field for the pseudospin of that rung. 
We assume that this field $h^z_T$ is proportional to the displacement $d_V$ 
of the V site of that rung and write:
\begin {equation}
h_{T}^z = 2g_{Is}d_{V}.
\end{equation}

The two subsystems are coupled due to the dependence of the coupling $J_{ij}$
on the charge distributions $\langle T_{ij}^z\rangle$ and  $\langle
T_{i+1j}^z\rangle$. $J_{ij}$ describes the coupling of spins $\vec{S}_{ij}$
and $\vec{S}_{i+1j}$ in one subsystem and this term is contained in one
subsystem, while $\langle T_{ij}^z\rangle$ and $\langle T_{i+1j}^z\rangle$ are
operators used in the description of the other subsystem. The coupling between
the charge-ordering evidence by nonzero $\langle T_{ij}^z\rangle$ and 
$\langle T_{i+1j}^z\rangle$ and the superexchange coupling $J_{ij}$
along the ladder is negative. With increasing charge ordering along a ladder
the effective superexchange coupling $J_{ij}$ decreases
as shown in Fig.~\ref{fig5}. This may change, though, if we include the 
effect of lattice distortions accompanying the charge ordering. We then have
to deal with a combination of effects on $J_{ij}$ as argued above. 

We first consider equation (\ref{eq2.10}) for each of the two subsystems
separately and then discuss the combined ordered ground-state. We can make a
Wigner-Jordan transformation of the spin operators of a subsystem to obtain
spinless fermions describing the spin degrees of freedom. Let $k_F=\pi/2b$ be
the Fermi momentum of these spinless fermions where $b$ is the lattice
constant in b direction. 

For a single subsystem, e.g., subsystem 1, alternation of the superexchange
$J_{ij}$ along the Heisenberg chains can be obtained by shifting the Na
ions alternatingly in c-direction or by shifting the O ions on the
rungs in b-direction as shown in Fig.~\ref{fig10}. Applying the result of 
Cross and Fisher~\cite{cross79} we arrive at renormalized
phonon frequencies for $q=2k_F$:
\begin {eqnarray}
\label {eq3.2}
\tilde{\omega}^2_{2k_F, O_R} &=& \omega^2_{2k_F, O_R}
-0.26\left|2g_{O_R}\left(2k_F\right)\right|^2 T^{-1}, \nonumber \\
\tilde{\omega}^2_{2k_F, Na}&=& \omega^2_{2k_F, Na}
-0.26\left|g_{Na}\left(2k_F\right)\right|^2 T^{-1}.
\end{eqnarray}
The coupling of a shift of the Na ion to a superexchange alternation 
is smaller
by a factor of 2, because a shift of the Na site in c-direction affects the
superexchange only on
one two-rung cluster whereas a shift of the rung O site in b-direction
affects the superexchange on two neighbouring two-rung clusters.

For a single subsystem, e.g, subsystem 1, there is also an
instability towards superexchange alternation along the Heisenberg chains 
accompanied by
charge ordering along the Ising chains. This corresponds to shifts 
induced by the lattice mode $q_0$ with $q_0=2k_F$ as shown in
Fig.~\ref{fig3}. Within this mode the V sites on the chains of the A
ladder shift in c-direction. The direction of the shift alternates along the
ladder, corresponding to a staggered longitudinal field for the pseudospins of
subsystem 1. 
The leg O ions on the B ladders also shift in c-direction together with the
neighbouring V sites of the A ladder. This causes an alternation of the
superexchange along the B ladder, corresponding to an alternation of $J_{ij}$
along the Heisenberg spin chains of subsystem 1. Applying the results of Cross
and Fisher and using RPA for the effect of the pseudospin-lattice coupling 
one obtains
\begin {equation} \label {eq3.3}
\tilde{\omega}^2_{q_0} = \omega^2_{q_0} -0.26\left|g_{VO}
\left(q_0\right)\right|^2
T^{-1} -4g_{Is}^2\chi_{q_0}\left(T\right)
\end {equation}
where the last term describes the influence of the Ising chains. 
Here $\chi_q\left(T\right)$ is the susceptibility of the Ising chain due to a
longitudinal field along z-direction. $g_{VO}$ describes the coupling of a
shift of the V sites on the superexchange of the neighbouring ladder
via a shift of the O site.

\subsection {The phase transition at $T_{C1}$}
\label {sec3.1}

We know from experiment that the phase transition at $T_{C1}$ in
$\alpha'$-NaV$_2$O$_5$ is accompanied by 
charge ordering~\cite{fagot00}. 
We therefore first consider equation (\ref{eq3.3}). We
can take $g$ for the $q_0$ mode in Fig.~\ref{fig3} directly
from the results in~\cite{bernert00}:
\begin {equation} \label {eq3.7}
\begin {array}{rcccl} 
g_{q_0}^{VO} &= & \sqrt{\frac{\hbar}{m_V}}\frac{\delta_J^{\mathrm{exp}} J}{d_V} &=& 7.884
\cdot 10^{9}~\mathrm{K~s}^{-1/2}
\end{array}
\end{equation}
with $d_V=6.0768$ pm being the square root of the average of the squared shifts
of the V sites on the A ladder, $\delta_J^{\mathrm{exp}}=0.26$ 
the exchange dimerisation 
for layer a, $J=522$ K and $m_V$ the V ion mass. We also need to obtain
$\chi_q(T)$. While this is very difficult for general values of $q$, it is
facilitated considerably when $q=q_0$. 
In this case the effective field for the pseudospins due to the
lattice distortion is staggered in b-direction. The Ising chain has an
antiferromagnetic coupling $K_{Lz}>0$ along b-direction. Calculating
$\chi_{q_0}(T)$ means 
therefore to calculate the susceptibility of an antiferromagnetic Ising chain
in a transverse field for the application of an infinitesimal 
staggered longitudinal field. This is equivalent
to calculating the susceptibility  
of a ferromagnetic Ising model in a transverse field for the application of an
infinitesimal constant longitudinal field. This susceptibility can be expressed
via the pseudospin correlation function $\rho(\lambda,T)$: 
\begin {equation} \label {eq3.8}
\chi_{q_0}\left(T\right) = \beta \lim_{N\to\infty}\frac{1}{N}\sum_{i,j=1}^{N}
\rho^z_{\left|i-j\right|}\left(\lambda,T\right)  
\end{equation}
where $\rho^z_{|i-j|}(\lambda,T)=\langle T_i^z
T_j^z\rangle(\lambda,T)$ and $\beta=1/k_BT$ is the inverse
temperature. $\lambda=\tilde{K}_{Lz}/4\tilde{t}_R$ is the ratio between
the pseudospin interaction and the transverse field and determines the
properties of the Ising chain. $\lambda=\infty$ corresponds to the Ising chain
without a transverse field, $\lambda=0$ to the paramagnetic limit.

For an Ising chain in a
transverse field $\rho^z_n(\lambda,T)$ can be written in the form of a
Toeplitz determinant~\cite{pfeuty70}. For large $n$ acccording to Szeg{\"o}'s
Theorem~\cite{grenander58} we have
\begin {equation} \label {eq3.9}
\rho^z_n\left(\lambda,T\right)=\frac{1}{4}P\left(\lambda,T\right)
G^n\left(\lambda,T\right) 
\end {equation}
where expressions for $P(\lambda,T)$ and $G(\lambda,T)$ 
are found in~\cite{mccoy68}. Some simple limits are given below.
To obtain $\chi_{q_0}(T)$ we assume that the asymptotic expression (\ref{eq3.9})
is correct for all $n$. With this assumption summation of the series
(\ref{eq3.8}) yields
\begin {equation} \label {eq3.10}
\chi_{q_0,\lambda} = \frac{1}{4}\beta P\left(\lambda,T\right)
\frac{1+G\left(\lambda,T\right)}{1-G\left(\lambda,T\right)}. 
\end{equation}
The approximation is expected to
be correct close to the quantum critical point, i.e., 
close to $\lambda=1$ at $T=0$, since then the susceptibility and
the correlation length diverge. However, we can obtain $P(\lambda,T)$ and
$G(\lambda,T)$  
for the case $\lambda=\infty$, i.e., the pure Ising chain, from
Ref.~\cite{mccoy68} as
\begin {equation} \label {eqA.5}
\begin {array} {rcl}
P_{\lambda=\infty}&=&1 \\
G_{\lambda=\infty} &=& \tanh\left(\beta\tilde{t}_R\lambda\right).
\end{array}
\end{equation}
When $\lambda=0$ we are in the paramagnetic regime and find
\begin {equation} \label {eqA.6}
\begin {array} {rcl}
P_{\lambda=0} &=& \frac{2}{2\beta\tilde{t}_R}
\left(1-\lambda^{2}\right)^{-3/4} = 
\frac{2}{2\beta\tilde{t}_R}, \\
G_{\lambda=0} &=& \lambda\left(\tanh\left[\beta\tilde{t}_R\left(1+\lambda\right)\right]\right)
\exp\left(\left(2\pi\beta\tilde{t}_R\right)^{-1/2}
\exp\left(-\beta\tilde{t}_R\left(1-\lambda\right)\right)\ldots\right) \\
&=& 0.
\end{array}
\end{equation}
With these expressions we can calculate $\chi_{q_0}(T)$ 
in the limits $\lambda=\infty$, $0$ and find
\begin {equation} \label {eq3.11}
\chi_{q_0,\lambda=\infty}\left(T\right) =
\frac{\beta}{4}\exp\left(\frac{1}{2}\beta\tilde{K}_{Lz}\right) 
\end {equation}
and 
\begin {equation} \label {eq3.12}
\chi_{q_0,\lambda=0}\left(T\right) = \frac{1}{2\left(2\tilde{t}_R\right)} \tanh
\left(\frac{1}{2}\beta\left(2\tilde{t}_R\right)\right).
\end {equation}
Both expressions (\ref{eq3.11}) and (\ref{eq3.12}) are equal to
the exact solution in these limits which
implies that this $\chi_{q_0,\lambda}$ 
should be a reasonable approximation for all values of
$\lambda$.  We can therefore in
principle calculate $T_{C1}$ from equation (\ref{eq3.3}) knowing $g_{Is}$,
$\omega_{q_0}$ and $\lambda$ by simply setting $\tilde{\omega}_{q_0}=0$.  

We can find $\omega_{q_0}$ from measurements of the elastic
constants~\cite{schwenk98}. There a strong anomaly in 
the $c_{66}$ mode of $\alpha'$-NaV$_2$O$_5$  was observed
at the phase transition temperature. This mode couples to a 
zig-zag like charge ordering and we therefore identify it with the $q_0$ mode
of the present 
calculations. Using the high temperature value for the sound velocity
$v_{66}=4200$ m/s we estimate $\omega_{q_0}=\pi v_{66}/b = 279$ K, where
$b=3.611$ {\AA} is the lattice constant in b-direction for the undistorted
lattice.

When $\omega_{q_0}$ is known we obtain $\lambda$ and $g_{Is}$ by fitting
them to $T_{C1}=34$ K and to the experimental value of the shift of $T_{C1}$ 
in a
magnetic field. The latter is reduced from its standard spin-Peierls
value~\cite{cross79b} due to the Ising part of the transition. This is seen as
follows. Considering only the spin-Peierls part of the transition,
applying a magnetic field
lowers the transition temperature $T_{C1}(H=0)$ without a magnetic field 
to a value $T_{C1}^{H,SP}$. On the other hand, the Ising chain
susceptibility $\chi_{q_0,\lambda}(T)$ increases 
with decreasing temperature. This causes an increase of the critical
temperature from  $T_{C1}^{H,SP}$ to the observed phase transition temperature
$T_{C1}(H)>T_{C1}^{H,SP}$ in a magnetic field. 

The expansion for $T_{C1}$ in a magnetic field $H$ for a pure spin-Peierls
system is~\cite{cross79b}:
\begin{equation} \label {eq3.16}
\frac{T_{C1}\left(H\right) - T_{C1}\left(H=0\right)}{T_{C1}\left(H=0\right)} =
-0.36 \left(\frac{\mu_B H}{k_B T_{C1}\left(H=0\right)}\right)^2.
\end {equation}

Due to the influence of the Ising part this expression is modified to
\begin {eqnarray} \label {eq3.17}
\frac{T_{C1}\left(H\right) - T_{C1}^H}{T_{C1}^{H,Is}\left(0\right)} &=& -0.36
\left(\frac{\mu_B H}{k_B T_{C1}^{H,Is}\left(0\right)}\right)^2 \nonumber \\
\frac{k_B T_{C1}^{H,Is}}{J}
\left(0\right) &=& 0.8 \frac{\left(g_{q_0}^{VO}\right)^2}{\pi
J\omega_{q_0,\mathrm{eff}}^2\left(T_{C1}\left(H\right)\right)}
\end{eqnarray}
in the present case with
\begin {equation}
\omega_{q_0,\mathrm{eff}}^2\left(T_{C1}\left(H\right)\right) =
\omega_{q_0}^2-4g_{Is}^2\chi_{q_0,\lambda}\left(T_{C1}\left(H\right)\right). 
\end {equation}
For small shifts this leads to
\begin{equation} \label {eq3.18a}
\frac{T_{C1}\left(H\right) - T_{C1}\left(0\right)}{T_{C1}\left(0\right)} = -\frac{0.36} {1+T_{C1}\left(0\right)a\left(T_{C1}\left(0\right)\right)}
\left(\frac{\mu_B H}{k_B T_{C1}\left(0\right)}\right)^2
\end {equation}
with
\begin{equation} \label {eq3.18b}
a\left(T\right) = -\frac{4g_{Is}^2
\frac{\mathrm{d}\chi_{q_0,\lambda}\left(T\right)} 
{\mathrm{d} T}}{\omega_{q_0}^2 - 4g_{Is}^2\chi_{q_0,\lambda}\left(T\right)}.
\end{equation}
From the experimental result for the shift of $T_{C1}=34$ K in a magnetic field
in~\cite{koppen98,bompadre00,schnelle99} we obtain $a\left(T_{C1}\right)
\approx 0.086$ K$^{-1}$.  
Fitting $g_{Is}$ and $\lambda$ to these two values we obtain
\begin {eqnarray} \label{eq3.19}
\lambda &=& 0.99850 \nonumber \\
g_{Is} &=& 1.688 \mathrm{~K/(pm~shift~of~V)}.
\end{eqnarray}
Note that this value for $\lambda$ is close to the value for the ratio 
$K_{IL}/K_{Lz}\approx 0.997$ from equation (\ref{eq2.7}). This indicates that
geometrical frustration plays an important role.
To see how stable the solution is towards a change of the parameters, we
evaluate the dependence of $T_{C1}$ on $\tilde{K}_{Lz}$, $\omega_{q_0}$ and
$g_{Is}$: 
\begin {eqnarray} \label {eq3.20}
\frac{\partial T_{C1}}{\partial \tilde{K}_{Lz}} &=& 1.06, \nonumber \\
\frac{\partial T_{C1}}{\partial \omega_{q_0}} &=& -0.14, \nonumber \\
\frac{\partial T_{C1}}{\partial g_{Is}} &=& 12 \mathrm{~pm}.
\end{eqnarray}
$T_{C1}$ is rather sensitive to small changes of $\tilde{K}_{Lz}$ or
$\lambda$. This sensitivity is due to the fact, that $(1-\lambda)\ll 1$. It
is much less and the stability therefore much improved compared to the result 
presented in~\cite{mostovoy00}, where $\lambda\approx
0.99986$ was suggested.

Next we include the coupling between the two subsystems. For this we will
use the mean-field values for $\langle T_{ij}^z\rangle,\langle
T_{i+1j}^z\rangle$. They are zero at $T_{C1}$ and therefore the  
critical temperature $T_{C1}$ remains unchanged in this approximation.
But a mean-field approximation is not reliable
close to the critical point of the transition where fluctuations must be taken
into account. We therefore consider the zero temperature free energy of
different ordered states. For this we need the dependence of the superexchange
$J^{ij}_0$ 
along a ladder on the charge ordering order parameter $\langle T_j^z\rangle$
on the same ladder. This is approximately
\begin {equation} \label {eq3.23}
J_{ij}^{\mathrm{eff}} = J^{ij}_0\left(1-4\left\langle
T_j^z\right\rangle^2\right) 
\end {equation}
where we neglect a small contribution to $J_{ij}^{\mathrm{eff}}$ 
from the diagonal hopping
in the case of complete charge ordering on the ladder, i.e., for $\langle
T_j^z\rangle=\frac{1}{2}$. As argued at the end of section~\ref{sec2}, 
$J_{ij}^{\mathrm{eff}}$ enters our equations in two
places. First, it effectively changes $\omega_{q_0}$ through
$J_0f\left(T_{ij}^z,T_{i+1j}^z\right)$. 
Since we used $\omega_{q_0}$ from experiment, this effect
is already included. Second, $J_{ij}^{\mathrm{eff}}$ modifies the energy gain
resulting 
from the spin degrees of freedom due to distortions and it is this effect we
have to consider now. Using the expression for the magnetic energy gain
from Ref.~\cite{barnes98} for the 
alternating Heisenberg chain we can write down the energy gain $dF(T=0)$ 
of the ground-state 
of the ordered system versus that of the ground-state of the disordered
system. The dimerisation parameter of the Heisenberg chains 
$\delta_{J}^{(1),(2)}$ with $\delta_{J}^{(1),(2)}=g_{q_0}^{VO}Q_{1,2}/J$ from
equation (\ref{eq3.7}) is the order parameter for each of the two
subsystems. $Q=d_V\sqrt{m_V/\hbar}$ is the canonical displacement. We find:
\begin {eqnarray} \label {eq3.24}
dF\left(T=0\right) 
&=& b\left(Q_1^2+Q_2^2\right) - E_{SP}\left(Q_1\right)\left(1-4\left\langle
T_2^z\right\rangle^2\right) \nonumber \\
&&-E_{SP}\left(Q_2\right)\left(1-4\left\langle
T_1^z\right\rangle^2\right) -2g_{Is}\left(\langle T_1^z\rangle
Q_1+\langle T_2^z\rangle Q_2\right) 
\end {eqnarray}
with
\begin {eqnarray} \label {eq3.25}
b&=& \frac{1}{2}\omega_{q_0}^2, \nonumber \\
E_{SP}\left(Q_{1,2}\right) &=& 0.3134 J \left(\delta_J^{(1),(2)}\right)^{4/3}, \nonumber \\ 
\langle T^z_{1,2}\rangle &=&
\frac{1}{2}\left(1-\exp\left(-4\chi_{Is}\left(T=0\right)g_{Is}Q_{1,2}
\right)\right).
\end{eqnarray}
Here we assumed that $T^z$ approaches $\frac{1}{2}$ exponentially if a
staggered parallel field $g_{Is}Q$ is applied. In equation (\ref{eq3.24}) 
$E_{SP}$ denotes the energy gain from the spin-Peierls
distortions, while the terms proportional to $g_{Is}$ 
denote the energy gain from charge ordering.
The first term stands for the elastic energy of the distortion.  
Optimizing equation (\ref{eq3.24}) in $\delta_J^{(1),(2)}\in[0,1]$ we
find a minimum at $\delta_{J}^{(1)}=0.023$, $\delta_Q^{(2)}=0$ 
or vice versa with
$dF=-0.80$ K. Requiring $\delta_{J}^{(1)}=\delta_{J}^{(2)}=\delta_J$, 
i.e., the equivalence of the
two subsystems we find $\delta_{J}=0.009$ with $dF=-0.35$ K. We have plotted
in Fig.~\ref{fig16} $dF$ for $\delta_{J}^{(1)}=\delta_{J}^{(2)}$ and
$\delta_{J}^{(2)}=0$. In the previous
calculations we used $\delta_J^{\mathrm{exp}}=0.26$ 
to obtain $g_{q_0}^{VO}$, so the first
result is more consistent with our calculation than the second one
and it also has the lower energy. 

From the derivation of $\chi_{Is}$ we can obtain the bare correlation length
along the Ising chains as
\begin {equation} \label {eq3.21}
\xi_{Is}\left(T\right) = -\frac{1}{\ln G\left(T\right)}.
\end{equation}
At $T_{C1}$ we find that $\xi_{Is}\approx 183$ lattice constants. This large
value of $\xi_{Is}$ 
might offer an explanation for the strong dependence of $T_{C1}$ 
on doping. By substituting Na with Ca or
depleting the system of Na additional electrons/holes are introduced
into the system. This implies double/zero occupancy of rungs. They weaken
correlations along 
the Ising chains. A rough estimate of the critical doping value at which the
phase transition temperature $T_{C1}$ is suppressed is $x_C\approx 0.5\%$. We
obtain it by assuming 
$x_C\xi_{Is}\approx 1$. This value is
in qualitative agreement with experiment. 
Substituting Na by Ca the transition disappears between $1\%$ and
$2.5\%$ of Ca~\cite{dischner00}. For hole doping due to Na deficiency the
transition disappears between 2\% and 3\% deficiency~\cite{isobe97}.

\subsection {The phase transition at $T_{C2}$}
\label {sec3.2}

\begin {figure}
\epsfig{file=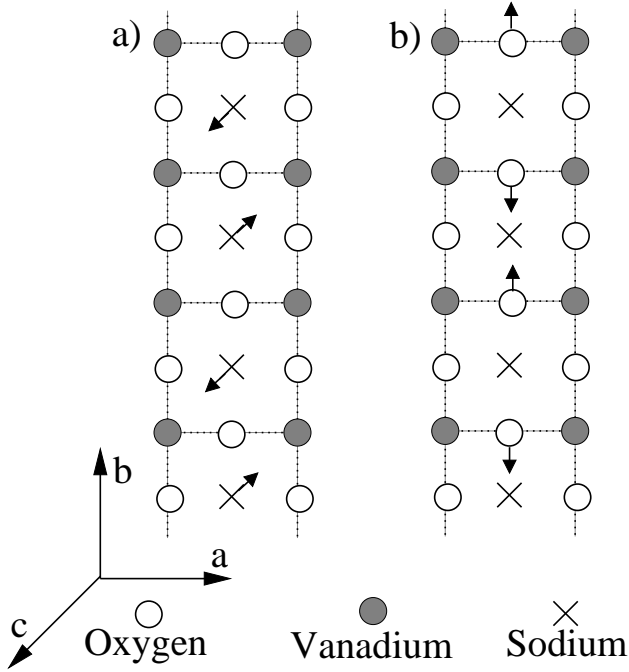}
\vspace{0.25cm}
\caption {a) Alternating Na shift in c-direction schematically shown
by arrows causes superexchange alternation. b) Alternating shift of the rung
O ions in b-direction causing superexchange alternation and inequivalence
of the Na sites along the ladder.}
\label {fig10}
\end{figure}

\begin {figure}
\epsfig{file=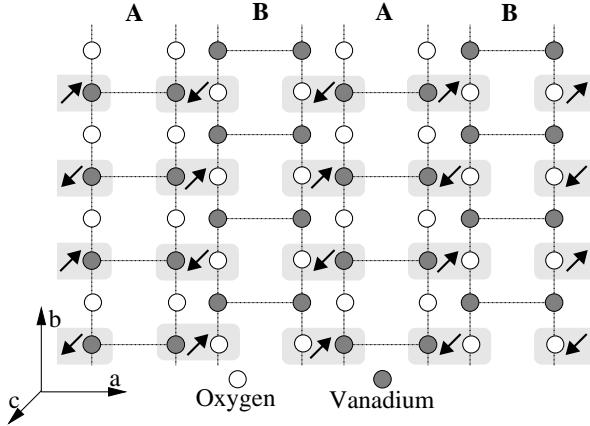}
\vspace{0.25cm}
\caption {The $q_0$ mode is shown, for which the 
V ion on one leg and O ion on neighbouring leg move
together in c-direction as shown by arrows.
This causes charge ordering on A ladders due to
a staggered change of the V on-site energies 
and superexchange alternation on B ladders due to O ion
motion.  In
experiment~\protect\cite{bernert00} shifts in a-direction are also observed
which are not shown here.}
\label {fig3}
\end{figure}

\begin {figure}
\epsfig{file=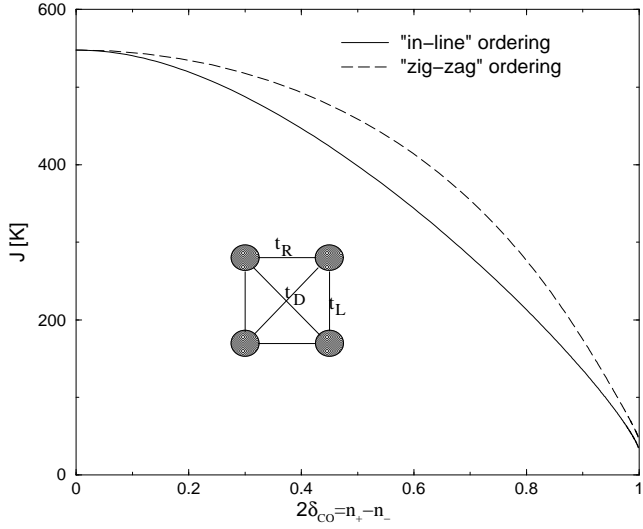,height=8.5cm, width=7.1cm, angle=-90}
\vspace{0.25cm}
\caption {Dependence of superexchange $J$ on charge ordering. Results from
cluster calculation for two rungs of the same ladder with
parameters $t_R=0.172$ eV, $t_L=0.049$ eV, $t_D=0.062$ eV, $V_L=0.344$ eV,
$V_R=0.398$ eV, $U=4$ eV and superexchange defined as the singlet-triplet
gap~\protect\cite{bernert00}. ``Zig-zag'' charge ordering or ``in-line'' 
charge ordering was induced by changing the on-site energies with
$\epsilon_1=\epsilon_3=-\epsilon_2=-\epsilon_4$ for ``in-line'' ordering and
$\epsilon_1=-\epsilon_2=-\epsilon_3=\epsilon_4$ for ``zig-zag''
ordering. Inset shows geometry of cluster. The difference in the behaviour of
the superexchange between the two
ordering patterns comes from the fact that $t_L\neq t_D$ and $V_L \neq V_D=0$.}
\label {fig5}
\end{figure}

\begin {figure}
\epsfig{file=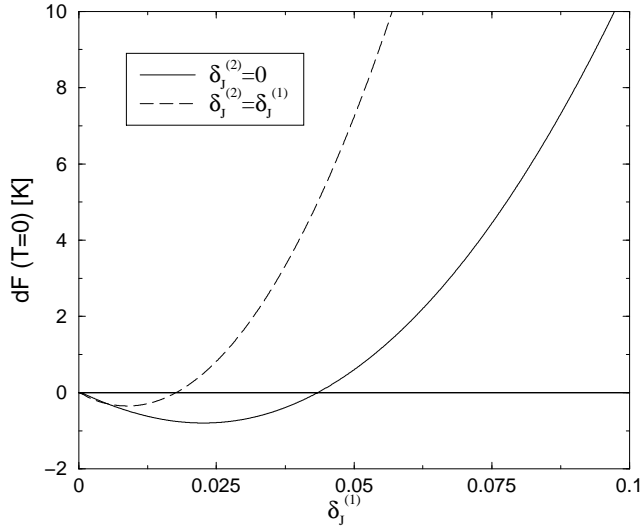,width=8.5cm,height=7.10cm,angle=0}
\caption {Ground-state energy gain $dF$ of the ordered system versus the disordered system 
depending on the order parameters.}
\vspace{0.25cm}
\label {fig16}
\end{figure}

Next we consider the second phase transition observed. According
to~\cite{fagot00} 
this transition is of the Ginzburg-Landau type. It opens a spin gap and
creates a local 
distortion at the Na sites as observed from the changes of the quadrupolar
electric field tensor. Charge ordering at the V sites starts
before this transition sets in~\cite{fagot00}. 
For these reasons and in order to explain the observation of eight
inequivalent Na sites in $^{23}$Na-NMR~\cite{fagot00,ohama00} 
we suggested in~\cite{bernert00}
that the opening of the spin gap in the system is due to an alternating shift
of the Na ions along the charge-ordered A ladders
as shown in Fig.~\ref{fig10}a. A second possibility is 
an alternating shift of the rung O ions on the A ladders as shown in
Fig.~\ref{fig10}b. 

By using the hopping matrix elements obtained in~\cite{bernert00} for the
charge-ordered ladder in layer a 
we can calculate the effective superexchange $J$
dependence on the charge ordering $\delta_{CO}$ on the same ladder. 
As we did for the undistorted ladders (see
Fig.~\ref{fig5}), the superexchange is defined by the
singlet-triplet gap on a two-rung cluster. It is obtained via
exact diagonalisation. We know from experiment~\cite{yosihama98} 
that $J \approx 440$ K in the low-temperature
phase. This value of $J$ corresponds to an incomplete charge ordering
$\delta_{CO}^0=\frac{1}{2}(n_+-n_-)\approx 0.32$
on the A ladders in agreement with experiment~\cite{fagot00}. 

The hopping matrix elements $t_L$, $t_D$, $t_R$ change due to the distortion
accompanying the charge ordering~\cite{bernert00}. We parametrize these
changes with
$t(\delta_{CO}=0)-t(\delta_{CO})\propto \delta_{lat}\propto\delta_{CO}$, and
normalize the prefactors such that the hopping matrix elements
from Ref.~\cite{bernert00} for the high
temperature undistorted ladder are obtained at $\delta_{CO}=0$ and those for
the charge ordered ladder are obtained at $\delta_{CO}=\delta_{CO}^0$.

\begin {figure}
\epsfig{file=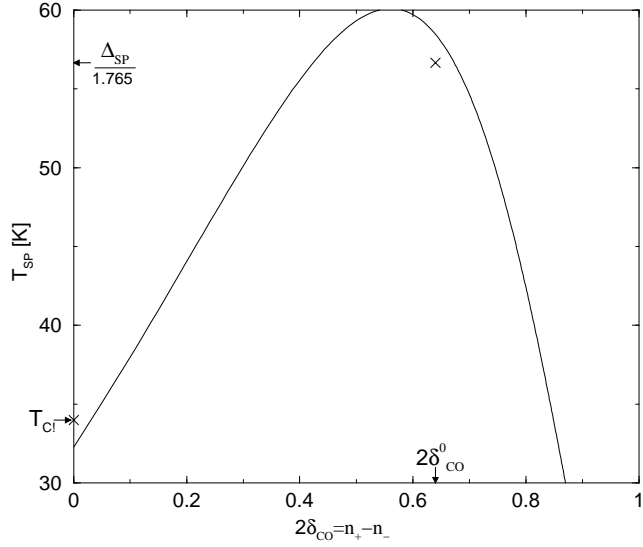,width=8.5cm,height=7.32cm,angle=0}
\vspace{0.25cm}
\caption {$T_{SP}$-dependence on charge ordering for a spin-Peierls
transition due to alternating shifts of the rung O ions on the A ladders in b
direction. Dimerisation of hopping matrix elements for a shift of 1 pm given by
$\delta_{t_D}=0.090$ and $\delta_{t_L}=-0.059$. $\times$ mark 
conditions (i) and (iii) (see text). The increase of $T_{SP}$ for small
$\delta_{CO}$  
is due to increase of the effective $\delta_J$ resulting from increase of
the hopping
matrix elements due to the distortion accompanying the charge ordering. 
The decrease for large $\delta_{CO}$ is caused by the decrease of the average
superexchange $\bar{J}$ as shown in Fig.~\ref{fig5}.}
\label {fig8}
\end{figure}

Due to alternating shifts of the rung O ions or the Na ions, the
effective hopping along the leg $t_L$ and the effective diagonal hopping $t_D$
alternate also. The size of this is given by the parameter 
$\delta_{t_{L,D}} = (t_+^{L,D}-t_-^{L,D})/(t_+^{L,D}+t_-^{L,D})$. 
If we know $\delta_{t_L}$ and $\delta_{t_D}$
we can find the superexchange alternation parameter $\delta_J$ 
by calculating the superexchange for the inequivalent two-rung
clusters~\cite{bernert00}. This 
gives us an approximate value for the spin-lattice coupling. Using it, the
parametrized $t_L$, $t_D$ and $t_R$ from above, and equation (\ref{eq3.2})
for the critical temperature we find $T_{SP}(\delta_{CO})$
for a spin-Peierls transition depending on
the charge ordering:
\begin {equation}
\frac{k_B T_{SP}\left(\delta_{CO}\right)}{J\left(\delta_{CO}\right)} = 0.8 
\frac{4g_{Ox}^2\left(\delta_{CO}\right)} {\pi J\left(\delta_{CO}\right)
\omega_{Ox}^2}.  
\end{equation}
We can estimate the value $\omega_{Ox}$ 
from the velocity $v_{22}$ of the $c_{22}$ mode in~\cite{schwenk98}. 
It describes a longitudinal mode along b-direction as required
for the proposed shifts of the rung O ions 
with $v_{22}=6500$ m/s which yields $\omega_{Ox} = 432$ K.
For given values of $\delta_{t_L}$ and $\delta_{t_D}$ we obtain the
dependence of the critical temperature $T_{SP}$ on the charge ordering
$\delta_{CO}$ as shown in Fig.~\ref{fig8}. In the low-temperature limit the
spin gap
is approximately given by $\Delta(\delta_{CO}^0)\approx 1.8
T_{SP}(\delta_{CO}^0)$. Therefore the BCS ratio is enhanced by
a factor of $T_{SP}(\delta_{CO}^0)/T_{C1}$. 

We search for values of $\delta_{t_L}$ and
$\delta_{t_D}$ for which the following conditions are fulfilled: (i)
without charge ordering the critical temperature for such a transition is
below 34 K, since the spin gap opens only below
the charge ordering temperature; 
(ii) with increasing
charge ordering the critical temperature is required to increase, because an
enhanced BCS ratio has been observed; (iii)
at $\delta_{CO}=\delta_{CO}^0$ the spin gap $\Delta(\delta_{CO})\approx 1.8
T_{C}^{SP}(\delta_{CO}^0)$ should be approximately 100 K as found
experimentally.  

The result is
found in Fig.~\ref{fig8}. The conditions (i) to (iii) are 
fulfilled for $\delta_{t_L}=-0.059$, $\delta_{t_D}=0.090$ and with
$T_{C1}=34$ K an effective BCS ratio of 6.05 is found. 
We therefore have a spin-Peierls phase
transition which is driven by charge ordering on the same ladder. 
When the
first transition takes place at about $T_{C1}=34$ K charge ordering sets in on
the A ladders. This changes the two-particle correlation functions and the
effective hoppings so that $T_{SP}$ increases as compared with its value
without charge ordering. At some temperature 
$T_{C2}< T_{C1}$ we have $T_{SP}(\delta_{CO}(T_{C2}))=T_{C2}$ and a 
spin-Peierls transition driven by charge ordering takes place. This is
caused by
an alternating shift of the O ions on the rungs of the A ladders,
probably together with a small alternating shift of the Na ions along the
A ladders. This
also causes an inequivalence of the Na sites along the A ladders, such
that we have 8 inequivalent Na sites as observed in
experiment~\cite{ohama00}. Since this happens before $\delta_{CO}$ attains its
final value, the spin gap at lower temperatures is larger than what would be
expected from the standard BCS ratio and $T_{C1}$.

In order to see whether the values for $\delta_{t_L}$ and $\delta_{t_D}$ are
plausible, we have investigated the effects of alternating 
shifts of the rung O or the
Na sites on the effective hopping matrix elements between V sites 
on the ladders. For this we use the Slater-Koster method as described
in~\cite{bernert00}. The results are given in Table~\ref{table3.1}. 
Those for $\delta_{t_L}$ and $\delta_{t_D}$ have been obtained for
otherwise undistorted ladders. $\delta_{t_L}$ and
$\delta_{t_D}$ are found to have opposite signs, as assumed above.
This can be understood by noticing that
an increase of diagonal hopping due to
increased hopping via the Na sites goes together
with a decrease of the hopping along the legs of the ladder. 
The effects of charge ordering
distortions and of the 
shifts of Na or rung O sites on the hopping matrix
elements are 
small. Therefore we may neglect higher-order effects resulting 
from combinations of the distortions. 

The results in the first three rows of 
Table~\ref{table3.1} are based on~\cite{bernert00} 
where we included V-O, V-V
and Na-O hopping matrix elements in the initial Hamiltonian,
projecting these on effective V-V hopping matrix
elements. However, it has been argued in~\cite{yaresko00} that direct hopping
between the O sites should also contribute to the effective
Hamiltonian. It remains unclear, which fraction of the total O-O hopping
matrix 
elements obtained in~\cite{yaresko00} results from direct hopping and which
fraction 
comes from indirect hopping via the neighbouring Na site. To investigate
the effect of such a hopping on the shifts of the rung
O ions we introduced O-O hopping matrix elements according to the
Slater-Koster approximation for into the initial Hamiltonian and calculated
effective V-V hopping matrix elements as before. One
finds $t_R=-0.17$ eV, $t_L=0.17$ eV, $t_D=-0.04$ eV, $t_{IL}=-0.16$
eV. Introducing an alternating shift of the rung O ions in b-direction by 1
pm leads to $\delta_{t_L}\approx 0.003$ and $\delta_{t_D}\approx 0.1$ as given
in the fourth row of Table~\ref{table3.1}. The
effect especially on the diagonal hopping is much stronger than it is without
inclusion of the direct O-O hopping matrix elements.
However, these values for $t_L$ and $t_{IL}$ are far from those found
by LDA~\cite{yaresko00} and ab-initio methods~\cite{suaud00}. One can therefore
conclude that at least within the Slater-Koster method the contribution of the
direct O-O hopping should be small as expected for next-nearest neighbour hopping, although it may help to explain
the small differences 
between the LDA results of~\cite{yaresko00} and our results
in~\cite{bernert00}. The results from
Table~\ref{table3.1} can serve only as an estimate of the 
values of the spin-lattice coupling. 

\section {Comparison with x-ray diffraction and neutron scattering
experiments} 
\label {sec4} 

In the following we
compare our results with experimental observations. We do this
with respect to two types of experiment: x-ray structure
determination~\cite{ludecke99,deboer00,bernert00}  
and inelastic neutron scattering results for the magnon
dispersion~\cite{yosihama98,grenier00}. We compute the latter for the
calculated structure by using a local-dimer approximation.

Comparing our results with x-ray structure determination we
find good agreement. In both cases we have a superstructure consisting of two
inequivalent ladder types alternating along $a$-direction: charge-ordered A
ladders and strongly dimerised B ladders. A phase with two A
ladders separated by a B ladder is also correctly
found. We can explain these findings with the analysis given in
section~\ref{sec3.1}: the displacement of the V sites on the A ladders and
of the O sites on the B ladders cause a combined
charge-ordering-spin-Peierls transition.

X-ray structure analysis finds  $Fmm2$ symmetry in the low-temperature
phase. In our model this symmetry is realized for $T_{C1}>T>T_{C2}$.
However, the phase transition at $T_{C2}$ 
obtained for our model 
in section~\ref{sec3.2} is accompanied by a breaking of this
symmetry. It results from the shift of the rung
O ions on the A ladders, as shown in Fig.~\ref{fig10}b. 
A low-temperature spin gap of $\Delta\approx
100$ K on these ladders requires a superexchange dimerisation of about 
$0.05$ when
$\Delta=2J\delta_J^{3/4}$ from Ref.~\cite{barnes98} is used with the
experimental value for $J=440$ K. 
The coupling constant at $\delta^0_{CO}=0.32$ is
approximately $0.043$ per pm shift of the rung O ion.
This corresponds to an actual shift of 1.1 pm of each rung O ion on
ladder A.

The intensity of the x-ray scattering for small scattering angles
is proportional to $Z^2$ where $Z$ is the ionic charge. Below $T_{C2}$ 
only one
O ion per two formula units shifts its position to break the  $Fmm2$
symmetry. In addition as estimated above 
these displacements are smaller than the displacements caused by the phase
transition at $T_{C1}$ and below, where a displacement of
7.46 pm and 4.26 pm for the V sites on the A ladder is found experimentally. 
It is therefore well possible that the
scattering peaks resulting from 
the lower symmetry cannot be observed within experimental
resolution. 
Insofar x-ray structure determination results do not contradict
our theoretical results for the second transition.

\begin {figure}
\epsfig{file=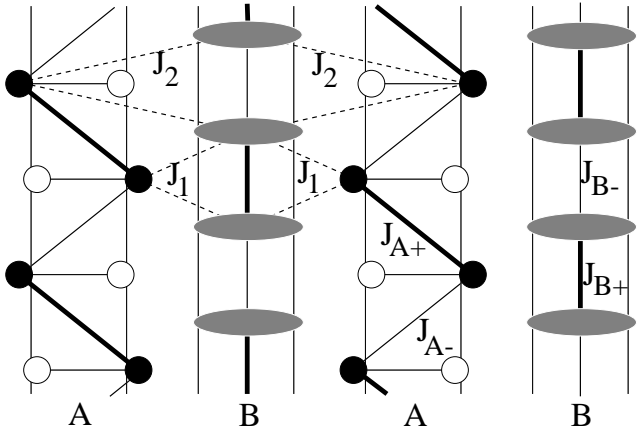}
\vspace{0.25cm}
\caption {Geometry of initial magnetic Hamiltonian
(\ref{eq4.1}). $J_{\pm}=J(1\pm\delta)$ for the respective ladder.}
\label {fig11}
\end{figure}

Magnon dispersion, however, should be greatly affected by a dimerisation.
The structure which is obtained after the two
transitions have taken place is schematically 
shown in Fig.~\ref{fig11}. The calculated magnon dispersion for this
structure should therefore be
a good test of the theory. For other suggested structures it was found that
the corresponding
magnon dispersion disagrees with experimental results: in the case of 
a lattice with ``zig-zag'' charge order on all ladders, the dispersion along
a-direction does not agree with experiments~\cite{thalmeier99}. 
For the spin-cluster
model proposed in~\cite{deboer00} 
the magnon dispersions, both in a- and b-direction,
do not agree with experiments either~\cite{trebst00,honecker00,gros00}. 

\begin {figure}
\epsfig{file=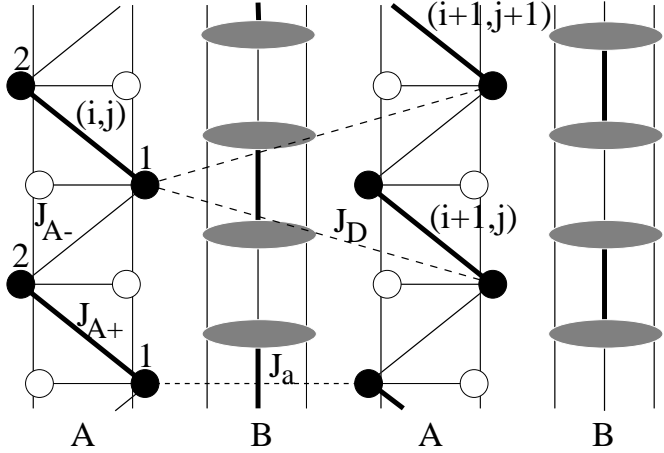}
\vspace{0.25cm}
\caption {Geometry of effective magnetic Hamiltonian (\ref{eq4.2}). 1 and 2
denote the two spins of a dimer, $J_a$ and $J_D$ denote the effective
superexchange between spins in the A ladder via the B ladder for nearest
interladder neighbours and next-nearest interladder neighbours. V$_{21}$ sites
with increased electron density denoted by black circles,  V$_{22}$ sites
with decreased electron density by white circles.} 
\label {fig12}
\end{figure}

\begin {figure}
\epsfig{file=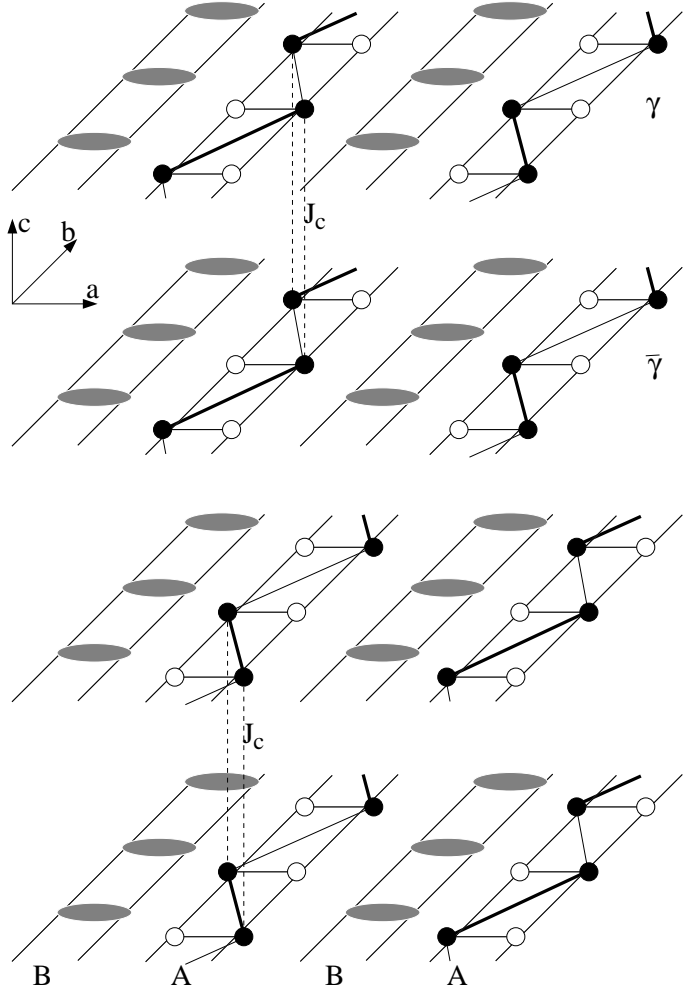}
\vspace{0.25cm}
\caption {Unit cell of low-temperature ordered state. Along the A ladders one
has pairs of V$_{21}$ sites (black circles) alternating with pairs of
V$_{22}$ sites (white circles) in c-direction. $J_c$ therefore connects two
layers $\gamma$, $\bar{\gamma}$ only. For such two layers, dimers on the A
ladders are 
assumed to lie directly above each other in c-direction.}
\label {fig13}
\end{figure}

We therefore calculate the magnon dispersion at a temperature well below
$T_{C2}$. 
In order to do this we use the spin-dimer representation
following Refs.~\cite{thalmeier99,leuenberger84,haley72}. 
The initial magnetic Hamiltonian is then
\begin {eqnarray} \label {eq4.1}
H &=& \sum_{i\in A} J_A\left(1+\left(-1\right)^j\delta_A\right)
\vec{S}_{ij}\vec{S}_{ij+1} \nonumber \\
&&
+\sum_{i\in B}
J_B\left(1+\left(-1\right)^j\delta_B\right) \vec{S}_{ij}\vec{S}_{ij+1}
\nonumber \\ 
&&
+\sum_{\langle ij,mn\rangle} J^{ij,mn}_{IL}\vec{S}_{ij}\vec{S}_{mn}
\end{eqnarray}
for a  geometry and $J^{ij,mn}_{IL}$ as shown in Fig.~\ref{fig11}. It
describes the pure spin part of equation (\ref{eq2.10}) 
with the effects of lattice distortions and pseudospins 
included within the effective coupling constants.
The large
value of $\delta_B\approx 0.26$ at $T=15$ K found in~\cite{bernert00} 
corresponds to a large gap in the excitation
spectrum of the B ladders. Using $\Delta_B=2J_B\delta_B^{3/4}$
for the spin gap~\cite{barnes98} and $J_B=52$ meV from Ref.~\cite{bernert00} 
we find $\Delta_B\approx 38$ meV. 
The B ladders should therefore not contribute directly
to the low-lying parts of the magnon dispersion. Since we are
only interested in these, we assume that we have an indirect exchange
coupling between A ladders through the virtual singlet-triplet excitations
of the B ladders. This results in an effective Hamiltonian:
\begin {eqnarray}\label {eq4.2}
H &=& \sum_{ij\gamma}J_A\left(1+\left(-1\right)^j\delta_A\right) \vec{S}_{ij\gamma}
\vec{S}_{ij+1\gamma} +\sum_{ij\gamma}J_a \vec{S}_{ij\gamma} \vec{S}_{i+1j\gamma} \nonumber \\
&&
+\sum_{ij\gamma}J_D \vec{S}_{ij\gamma}
\left(\vec{S}_{i+1j+1\gamma}+\vec{S}_{i+1j-1\gamma}\right) 
\nonumber \\
&&
+\sum_{ij\gamma} J_c\vec{S}_{ij\gamma}\vec{S}_{ij\bar{\gamma}}
\end{eqnarray}
with corresponding geometry and couplings as shown in Fig.~\ref{fig12}. This
exchange Hamiltonian exhibits manifestly a doubling of the period along the
a axis. Here
we also introduced a superexchange $J_c$ in c-direction between layers; it
is assumed to exist mainly between the V$_{21}$ sites which lie
directly above each other~\cite{bernert00}. We therefore have $\gamma=\pm 1$: 
Along c-direction the V sites of the A ladder are ordered as
\ldots--V$_{21}$--V$_{21}$--V$_{22}$--V$_{22}$--\ldots~as 
shown in Fig.~\ref{fig13}~\cite
{ludecke99,deboer00,bernert00}. Next we  
introduce dimer variables. We denote each dimer within the A ladder (see 
Fig.~\ref{fig12}) by coordinates $ij\gamma$, 
which are not identical with the former coordinates of the spins. As
denoted in Fig.~\ref{fig12} we then have spins $\vec{S}_{ij\gamma1,2}$. The
dimer 
variables are then:
\begin {equation} \label {eq4.3}
\begin {array}{rcl}
\vec{K}_{ij\gamma} &=& \vec{S}_{ij\gamma1} +  \vec{S}_{ij\gamma2} \\
\vec{L}_{ij\gamma} &=& \vec{S}_{ij\gamma1} -  \vec{S}_{ij\gamma2}.
\end{array}
\end{equation}
With those, equation (\ref{eq4.2}) takes the form
\begin {eqnarray} \label{eq4.4}
H &=& \frac{1}{4} J_A\left(1+\delta_A\right) \sum_{ij\gamma}
\vec{K}_{ij\gamma}\vec{K}_{ij\gamma}-\vec{L}_{ij\gamma}\vec{L}_{ij\gamma} 
\nonumber \\
&&
+\frac{1}{4} J_A\left(1-\delta_A\right) \sum_{ij\gamma}
\vec{K}_{ij\gamma}\vec{K}_{ij+1\gamma}-\vec{L}_{ij\gamma}\vec{L}_{ij+1\gamma} 
\nonumber \\
&&
+\frac{1}{4} J_a \sum_{ij\gamma}
\vec{K}_{ij\gamma}\vec{K}_{i+1j\gamma}-\vec{L}_{ij\gamma}\vec{L}_{i+1j\gamma} 
\nonumber \\
&&
+\frac{1}{4} J_D \sum_{ij\gamma} \left(\vec{K}_{ij\gamma}-\vec{L}_{ij\gamma}\right) \left(
\vec{K}_{i+1j\gamma}-\vec{L}_{i+1j\gamma}\right.\nonumber \\
&&
\left. \qquad +\vec{K}_{i+1j+1\gamma}-\vec{L}_{i+1j+1\gamma}\right) 
+\left(\vec{K}_{ij\gamma}+\vec{L}_{ij\gamma}\right) 
\nonumber \\
&&
\qquad\times\left(
\vec{K}_{i+1j\gamma}+\vec{L}_{i+1j\gamma}+\vec{K}_{i+1j+1\gamma}+\vec{L}_{i+1j+1\gamma}\right)
\nonumber \\
&&
+\frac{1}{2} J_c \sum_{ij\gamma} \vec{K}_{ij\gamma}\vec{K}_{ij\bar{\gamma}}+
\vec{L}_{ij\gamma}\vec{L}_{ij\bar{\gamma}}.
\end{eqnarray}
As in Ref.~\cite{thalmeier99} we will only use the products
$L_{ij\gamma}L_{mn\gamma'}$, since the 
other terms do not contribute to the dispersion of the spin excitations. We
then transform the $\vec{L}$ to
\begin {equation}\label {eq4.5}
\begin {array}{rcl}
\vec{L}_{ij\gamma} &=&
\frac{1}{\sqrt{2}}\left(\vec{M}^+_{ij}+\vec{M}^-_{ij}\right)\\ 
\vec{L}_{ij\bar{\gamma}} &=&
\frac{1}{\sqrt{2}}\left(\vec{M}^+_{ij}-\vec{M}^-_{ij}\right). 
\end{array}
\end{equation}
Furthermore we assume that the ladder designated by $(i+1,\gamma)$ 
is equivalent to the
ladder $(i,\bar{\gamma})$. This corresponds to a geometry of the A
ladders shown in Fig.~\ref{fig13}. We therefore set
$L_{ij\gamma}=L_{i+1j\bar{\gamma}}$, and separate the Hamiltonian into two
parts: 
\begin {eqnarray}\label{eq4.6}
H_{M^+} &=& -\frac{1}{4}\left(J_A\left(1+\delta_A\right)-J_c\right) 
\vec{M}^+_{ij}\vec{M}^+_{ij} 
\nonumber \\&&
- \frac{1}{4}J_A\left(1-\delta_A\right)\vec{M}^+_{ij}\vec{M}^+_{ij+1} 
-\frac{1}{4}J_a\vec{M}^+_{ij}\vec{M}^+_{i+1j} 
\nonumber \\&&
+\frac{1}{2}J_D\left(\vec{M}^+_{ij}\vec{M}^+_{i+1j}
+\vec{M}^+_{ij}\vec{M}^+_{i+1j+1}\right),\nonumber \\
H_{M^-} &=& -\frac{1}{4}\left(J_A\left(1+\delta_A\right)+J_c\right) 
\vec{M}^-_{ij}\vec{M}^-_{ij} 
\nonumber \\&&
- \frac{1}{4}J_A\left(1-\delta_A\right)\vec{M}^-_{ij}\vec{M}^-_{ij+1} 
+\frac{1}{4}J_a\vec{M}^-_{ij}\vec{M}^-_{i+1j} 
\nonumber \\&&
-\frac{1}{2}J_D\left(\vec{M}^-_{ij}\vec{M}^-_{i+1j}
+\vec{M}^-_{ij}\vec{M}^-_{i+1j+1}\right).
\end{eqnarray}

\begin {figure}
\epsfig{file=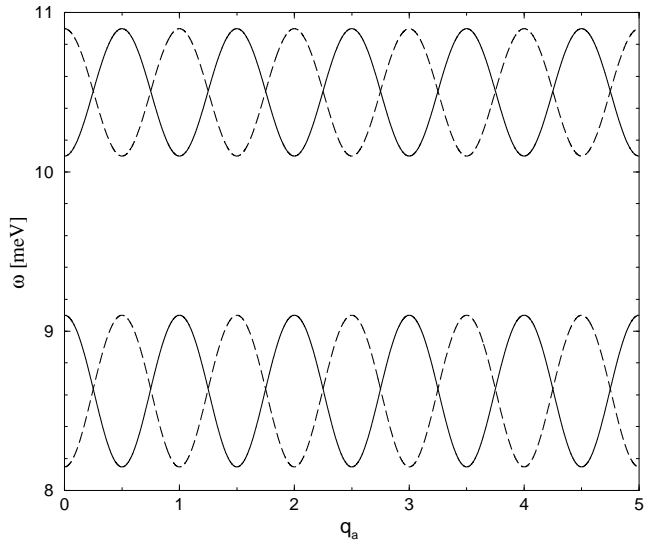,height=8.5cm, width=7.27cm,angle=-90}
\vspace{0.25cm}
\caption {Magnon dispersion along a-direction for $q_b=\frac{1}{2}=Q_b^{AF}$
(solid lines) and $q_b=1=Q_b^{ZC}$ (dashed lines).}
\label {fig14}
\end{figure}

The first term of each Hamiltonian describes an effective 
dimer with interaction strength
$J_A\left(1+\delta_A\right)\pm J_c$. Its dynamical
susceptibility is
\begin {equation} \label {eq4.7}
u_{\alpha\beta}^\pm\left(\omega\right) =
\delta_{\alpha\beta}\left(\delta_{\alpha x}+\delta_{\alpha y}\right)
\frac{2\left(J_A\left(1+\delta_A\right)\mp J_c\right)}
{\left(J_A\left(1+\delta_A\right)\mp J_c\right)^2 -\omega^2}.
\end{equation}
The dynamical RPA susceptibility of coupled dimers without anisotropy
of the superexchange is given by
\begin {equation} \label {eq4.8}
{\chi}\left(\vec{q},\omega\right) = \left[ 
1 -
{J}\left(\vec{q}\right) 
{u_{xx}}\left(\omega\right)\right]^{-1}
{u_{xx}}\left(\omega\right)
\end{equation}
where $J$ is the exchange term between dimers as given by the last four terms
of $H_{M+}$ and $H_{M-}$. After Fourier transformation these,
$J^{+}$ for $H_{M+}$ and $J^{-}$ for $H_{M-}$ are given by
\begin {eqnarray} \label {eq4.9}
J^{\pm}\left(\vec{q}\right) 
&=& \frac{1}{2}J_A\left(1-\delta_A\right) \cos 2k_y \nonumber \\
&&
\pm
\left(\frac{1}{2}J_a\cos\left(k_x-k_y\right) -2J_D\cos k_x\cos k_y\right).
\end{eqnarray}
Here $k_x=2\pi q_x/a$, $k_y=2\pi q_y/b$ with $a$ and $b$ being the lattice
constants of the high temperature unit cell. We note a
doubling of the unit cell in b-direction as required before
by equation (\ref{eq4.2}). The spin-wave excitations are obtained from the
poles of   
${\chi}\left(\vec{q},\omega\right)$. With equations (\ref{eq4.7}) and
(\ref{eq4.9}) we obtain
\begin {eqnarray} \label {eq4.10}
\omega^2_\pm &=& \left(J_A\left(1+\delta_A\right)\mp J_c\right)^2 - 
 \left(J_A\left(1+\delta_A\right)\mp
 J_c\right) \nonumber \\
&&
\times\left(J_A\left(1-\delta_A\right) \cos 2k_y \pm 
\left(J_a\cos\left(k_x-k_y\right) \right.\right. \nonumber \\
&&
\qquad\qquad\qquad\qquad \left.\left. -4J_D\cos k_x\cos k_y\right)\right).
\end{eqnarray}
We set $J_a^{\mathrm{eff}}=J_a-4J_D$ and $J_A=440$ K, as
used in our 
previous analysis and obtained from experiment~\cite{yosihama98}. Then we fit
the unknown parameters $J_a^{\mathrm{eff}}$, $J_c$ and $\delta_A$ to the
experimental values for three of the four gaps obtained from $\omega_+$ and
$\omega_-$. These gap values have been observed for $T\le 4.2$ K at $q_x=0$,
$q_y=\frac{1}{2}$ and $q_x=0$, $q_y=1$~\cite{grenier00}. 
Fitting three of the four gaps of sizes 10.9 meV,
10.1 meV, and 9.1 meV we obtain
\begin {equation} \label{eq4.11}
\begin {array} {rcl}
J_a^{\mathrm{eff}} &=& 0.212\mathrm{~meV},\\
J_c &=& 0.431\mathrm{~meV},\\
\delta_A &=& 0.0310.
\end{array}
\end{equation}
This determines 
the fourth gap to be 8.14 meV which is in excellent agreement with
the experimental value 8.2 meV from Ref.~\cite{grenier00}. 
In the model presented here
these spin gaps result from the assumed dimerisation of the charge
ordered A ladders whereas previously the origin of the
spin gaps was suggested to lie in an anisotropy of the
superexchange~\cite{thalmeier99}. 
The dispersion of $\omega^{\pm}$ along $a$-direction for
$q_y=\frac{1}{2},1$ are shown in Fig.~\ref{fig14}. These curves 
also agree with experiments. 

The low value of $J_a^{\mathrm{eff}}$, which causes the dispersion along
a-direction is due to
an effective coupling of next-nearest-neighbour
ladders. Furthermore, $J_a$ is partly compensated 
due to $J_D$ couplings. 

The value
for $J_c$, which causes the gap between the two modes, 
corresponds to a hopping $t_c\approx 0.021$ eV between neighbouring 
V sites of neighbouring layers when $J_c=\frac{4t_c^2}{U}$
and $U=4$ eV are used. 
This is in reasonable agreement with the value $t_c=0.015$ eV 
found in~\cite{bernert00} from the low-temperature structural data
by a Slater-Koster approximation.

\begin {figure}
\epsfig{file=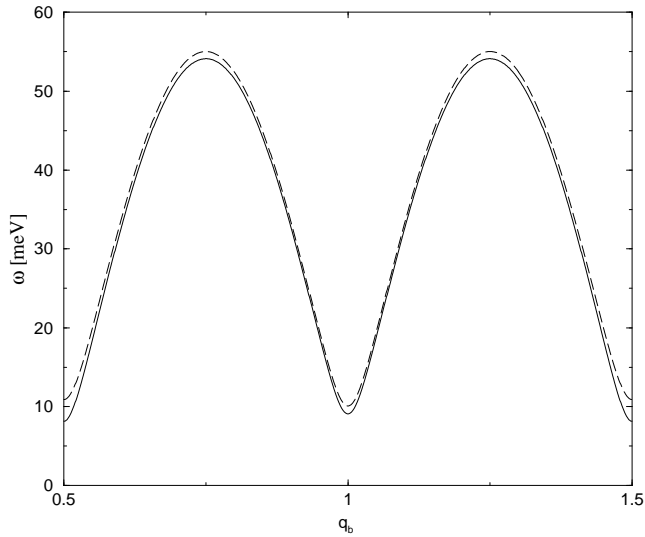,height=8.5cm, width=7.17cm,angle=-90}
\vspace{0.25cm}
\caption {Magnon dispersion along b-direction for $q_a=3.5$.} 
\label {fig15}
\end{figure}

We can also obtain the approximate dispersion in $b$-direction for $q_x=0$ as
shown in Fig.~\ref{fig15}. The maximum is at $\omega\approx 55$ meV and
agrees nicely with the value $\omega_{max}=59.5$ meV estimated
in~\cite{yosihama98} 
from experimental data. 

We therefore conclude that the results of our analysis in section \ref{sec3}
agree well with both the experimental results of x-ray structure determination
and the magnon dispersion found by inelastic neutron scattering results for
the magnon dispersion. 

\section {Summary and conclusions}

In this article we provided a theoretical description of the phase transitions
in $\alpha'$-NaV$_2$O$_5$. In section~\ref{sec2} we started from a
single band extended Hubbard model describing the hopping of electrons between
V sites. We projected this model onto an effective spin-pseudospin
Hamiltonian similar to that in Refs.~\cite{thalmeier98,sa00,mostovoy00}.  
Using parameters obtained from a previous Slater-Koster
analysis and findings of LDA+U
we argued that within the
given model a phase transition cannot occur from Coulomb interactions only: 
the effective lattice for such a transition is triangular and its geometric
frustration suppresses charge order, a distinctive feature which is present in
the observed transition. 
We therefore constructed a minimal microscopic model
based on a spin-pseudospin Hamiltonian which incorporates the 
coupling between charges, spins and the lattice. 

In section~\ref{sec3} we analyzed this model with regard to phase
transitions. We divided the model into two subsystems. Each of it
describes the
charge degrees of freedom on the rungs of half the ladders and the spin
degrees of freedom on the other ladders.
For each subsystem we
find an instability towards a transition at $T_{C1}$
which causes ``zig-zag'' charge ordering  
and superexchange dimerisation. With
this combined spin-Peierls-Ising transition we can explain the anomalous shift
of $T_{C1}$ 
in a magnetic field. It is reduced from its normal spin-Peierls value
due to the influence of the charge ordering. By calculating the free energy at
$T=0$ we found that the system should enter a phase where 
half of the ladders are ``zig-zag''
charge-ordered (A ladders), while the other half (B ladders) 
shows a strong superexchange alternation. The reason for this asymmetry lies in
a competition between 
the order parameters for charge ordering and superexchange dimerisation due to 
a shift of the leg O sites on the same ladder. This 
explains observations of x-ray structure determination. Due to the strong
effect of the one-dimensional Ising chains we can also
explain qualitatively 
the strong dependence of the transition on depletion of Na or 
substitution of Na with Ca.

By analyzing the coupling between superexchange dimerisation and lattice
distortion in the charge-ordered ladders 
we found that
another transition is induced. Charge ordering increases the coupling between
an alternating shift of the O sites on a rung and the superexchange
alternation beyond the threshold value of a second transition at 
$T_{C2}$. The system then enters a phase where the
charge-ordered A ladders dimerise and a spin gap opens.
Due to the alternation the Na sites along the A
ladders become inequivalent, such that we have 8 inequivalent Na sites as
observed in $^{23}$Na-NMR~\cite{fagot00,ohama00}, not only 6 as would be the case
in the $Fmm2$ symmetry indicated by x-ray structure determination. 

This also helps to explain the discrepancy between measurements of the critical
exponent of the lattice distortion~\cite{ravy99,nakao99,gaulin00} 
and of the critical exponent of
the spin 
gap opening~\cite{fertey98}: the former is smaller than the latter whereas one
would expect the opposite for a single transition due to $\Delta\propto
\delta_{lat}^{3/4}$. From the observations of  a
logarithmic peak in the  specific heat at $T_{C1}$ and
the observation of fluctuations by x-ray diffuse
scattering~\cite{gaulin00} we conclude that the transition at $T_{C1}$
is mainly of
2D Ising character whereas the second transition at $T_{C2}$, 
which opens the spin gap, can be described within a mean-field theory.
Since the charge ordering is not complete, when the second phase transition is
triggered, the coupling constant further increases for decreasing 
temperatures. This leads to an additional 
increase of the spin gap and also of the
BCS ratio. The conventional 
BCS ratio does not account for the
temperature dependence of the coupling constant $g_{q_0}^{VO}$. 

Within experimental resolution we find agreement with the x-ray structure
determination. The second transition breaking the $Fmm2$ symmetry
comprises only a small shift of every tenth O
site. We also find excellent agreement with experimental results for
the magnon dispersion at a temperature well below $T_{C2}$.

In conclusion we presented a microscopic model for $\alpha'$-NaV$_2$O$_5$
which yields two 
phase transitions close to each other. The first is a ``spin-Peierls-Ising''
transition causing charge order and superexchange dimerisation. The second is
a pure spin-Peierls transition triggered by an increase of the coupling
constant due to charge ordering.
With this model we can explain qualitatively and quantitatively a number of
experimental observations, i.e., the existence of two transitions, the general
structure of the 
low-temperature phase, the anomalous shift of $T_{C1}$ in a magnetic field,
the anomalous BCS ratio, the strong
dependence of the transition on doping, and the observed low-energy magnon
dispersion.

\bibliographystyle{prsty}
\bibliography {bibliothek}

\clearpage

\begin {table}
\begin {tabular} {l|ccc}
& $\delta_{t_D}=\frac{t_{D+}-t_{D-}}{t_{D+}+t_{D-}}$ 
& $\delta_{t_L}$ 
& $\delta_{J}=\frac{J_{+}-J_{-}}{J_{+}+J_+}$ \\
\hline
Na shift in\\
c-direction & 0.0105 & -0.0135 & 0.0035\\
V shift in\\
b-direction & 0.0236 & 0.0136 & 0.0347\\
Rung O shift \\
in b-direction & 0.0072 & -0.0093 & -0.0016\\
Rung O shift \\
in b-direction with \\
$t_{OO}$ included & 0.102 & 0.003 & 0.096
\end{tabular}
\vspace{0.25cm}
\caption {Change of hopping matrix elements for an alternating Na shift in
c-direction, an alternating V shift in b-direction and an 
alternating shift of rung O ions in b-direction
respectively. Shift is by 1 pm per site on an otherwise undistorted
ladder. The accompanying change of the superexchange has been calculated as
the singlet-triplet gap on a cluster of two neighbouring rungs of a
ladder~\protect\cite{bernert00}.}
\label {table3.1}
\end{table}

\clearpage

\clearpage

%
%

%
%

\end{document}